\documentclass[pdftex]{pasj01}
\Received{$\langle$reception date$\rangle$}
\Accepted{$\langle$acception date$\rangle$}
\Published{$\langle$publication date$\rangle$}

\usepackage{bm}	
\begin{document}
\title{ Impact of Turbulent Magnetic Fields on Disk Formation and Fragmentation in First Star Formation}
\author{Kenji Eric Sadanari\altaffilmark{1,*}}%
\author{Kazuyuki Omukai\altaffilmark{2}}%
\author{Kazuyuki Sugimura\altaffilmark{3}}%
\author{Tomoaki Matsumoto\altaffilmark{4}}%
\author{Kengo Tomida\altaffilmark{2}}%
\altaffiltext{1}{Department of Physics, Konan University, Okamoto, Kobe, Hyogo 658-8501, Japan}
\altaffiltext{2}{Astronomical Institute, Tohoku University, Aoba, Sendai, Miyagi 980-8578, Japan}
\altaffiltext{3}{Faculty of Science, Hokkaido University, Sapporo, Hokkaido 060-0810, Japan}
\altaffiltext{4}{Faculty of Sustainability Studies, Hosei University, Fujimi, Chiyoda, Tokyo 102-8160, Japan}
\email{k.sadanari@konan-u.ac.jp}

\KeyWords{stars:formation, Population$\rm I\hspace{-.1em}I\hspace{-.1em}I$, stars:magnetic field}

\maketitle

\begin{abstract}
Recent cosmological hydrodynamic simulations have suggested that the first stars in the universe often form as binary 
or multiple systems. However, previous studies typically overlooked the potential influence of magnetic fields during this 
process, assuming them to be weak and minimally impactful. Emerging theoretical investigations, however, propose an 
alternative perspective, suggesting that turbulent dynamo effects within first-star forming clouds can generate strong 
magnetic fields.
In this study, we perform three-dimensional ideal magnetohydrodynamics simulations, starting from the gravitational collapse of a turbulent 
cloud core to the early accretion phase, where disk fragmentation frequently occurs. Our findings reveal that turbulent 
magnetic fields, if they reach an equipartition level with turbulence energy across all scales during the collapse phase, 
can significantly affect the properties of the multiple systems. Specifically, both magnetic pressure and torques 
contribute to disk stabilization, leading to a reduction in the number of fragments, particularly for low-mass stars.
Additionally, our observations indicate the launching of protostellar jets driven by magnetic pressure of toroidal fields, although 
their overall impact on star formation dynamics appears to be minor. Given the case with which seed magnetic fields 
amplify to the full equipartition level, our results suggest that magnetic fields likely play a significant role in shaping the 
initial mass function of the first stars, highlighting the importance of magnetic effects on star formation in the early 
universe.
\end{abstract}

\section{Introduction}\label{Sec:introduction}
The first stars in the universe, formed from primordial hydrogen and helium gas and commonly 
referred to as Population III (Pop III) stars or metal-free stars, are believed to have 
emerged in the Universe around redshift $z\sim 10-30$ according to the standard cosmology. 
Their appearance marks a significant change in the cosmic landscape as they initiated the 
process of reionizing the Universe (e.g., \cite{Barkana_Loeb2001}). When these stars reached 
the end of their lives and exploded as supernovae (SN), heavy elements synthesized within 
them were released into the interstellar medium, altering the thermal properties of the gas 
and triggering the transition to the next generation of star formation, namely Pop II/I 
stars (e.g., \cite{Heger2003}; \cite{Umeda_Nommoto2003}; \cite{Greif2010}; \cite{Chiaki2018}). 

Studies on the formation of the first stars have been actively pursued since the development 
of the Big Bang Cosmology in the 1960s. Over the past few decades, significant advancements 
have been made in our theoretical understanding. According to the current understanding, 
the formation of the first stars occurs within so-called minihalos, which are dark matter (DM) 
halos with masses ranging from $10^{5} - 10^{6} M_{\odot}$ 
(e.g., \cite{Couchman_Rees1986}; \cite{Yoshida2003}; \cite{Greif2015}). 
The star formation is triggered by $\rm H_{2}$ cooling, which enables the baryonic gas to cool, 
leading to gravitational collapse. In the course of this, the cooling efficiency decreases as 
the $\rm H_2$ rotational levels reach the local thermodynamic equilibrium at densities of 
around $10^4\ \rm cm^{-3}$, causing the gas to enter a stage known as the ``loitering stage'' (e.g., \cite{Bromm2002}), characterized by slow contraction, before eventually forming a cloud core in a state of quasi-hydrostatic equilibrium. As this cloud core grows and approaches the Jeans mass ($M_{\rm J} \sim 10^{3}\ M_{\odot}$) through gas accretion, it undergoes runaway collapse, leading to the birth of protostars at its center (e.g., \cite{Abel2002}; \cite{Bromm2002}; \cite{Yoshida2008}).

After the formation of protostars, they undergo a phase of mass growth through gas accretion from the surrounding envelope, known as the accretion phase. Due to the high temperatures typically on the order of a few hundred K, the accretion rate can reach around $\dot{M}\sim 10^{-3}\ \rm M_{\rm \odot}\ \rm yr^{-1}$ 
(e.g., \cite{Stahler1986}; \cite{Omukai_Nishi1998}), significantly higher than in the case of present-day star formation. With such high accretion rates, the radiative feedback from the stars becomes inefficient, potentially allowing the first stars to grow to masses exceeding a few tens to even a few hundred solar masses if we assume minimal fragmentation and the formation of a single protostar (e.g., \cite{Omukai_Palla2003}; \cite{McKee_Tan2008}; \cite{Hosokawa2016})

However, a number of hydrodynamical simulations (e.g., \cite{Clark2011}; \cite{Smith2011}; \cite{Greif2012}; \cite{Stacy_Bromm2013}; \cite{Susa2019}; \cite{Chon_Hosokawa2019}; \cite{Kimura2021}; \cite{Park2021}; \cite{Sugimura2020}, \yearcite{Sugimura2023}) 
have revealed that disk fragmentation due to gravitational instability tends to occur under conditions of high accretion rates. \citet{Susa2019} have further shown that the number of fragments increases over time, resulting in highly multiple systems containing both massive and low-mass stars, with some even less than the solar mass. In these systems, low-mass stars can be easily ejected from the disk due to gravitational interactions within few-body systems (e.g., \cite{Clark2008}; \cite{Greif2012}; \cite{Susa2014}). While low-mass first stars theoretically should still survive until the present-day Universe, they have yet to be detected despite concerted observational efforts 
(e.g., \cite{Hartwig2015}; \cite{Ishiyama2016}; \cite{Magg2018} ). Moreover, massive binary systems evolving into binary black holes (BHs) could potentially serve as sources of gravitational wave events (e.g., \cite{Kinugawa2014}, \yearcite{Kinugawa2016}; \cite{Abbott2016}; \cite{Hartwig2016}). However, current simulations struggle to reproduce close binaries capable of merging within the age of the Universe, as accretion of high angular momentum gas causes the binary separation to expand to more than $10^{3}\ \rm au$ (e.g., \cite{Sugimura2020}, \yearcite{Sugimura2023}; \cite{Park2024}).
\begin{table*}
 \caption{Initial parameters of simulated cloud core}
 \label{table_cloud}
 \centering
  \begin{tabular}{lc}
   \hline \hline
   Parameter  &  Value\\
   \hline 
   mass $M_{\rm cl}$               & $5.5\times10^{3}\ M_{\odot}$\\
   radius $R_{\rm cl}$             &  $1.1\times10^{6}\ \rm au$\\
   central number density $n_{\rm c,0}$  & $1.4\times10^{3}\ \rm cm^{-3}$\\
   temperature $T_{\rm init}$       & $198\ \rm K$\\
   ratio of thermal to gravitational energies $E_{\rm th}/|E_{\rm g}|$ & $0.6$\\
   ratio of rotational to gravitational energies $E_{\rm rot}/|E_{\rm g}|$ & $0.01$\\
   ratio of turbulent to gravitational energies $E_{\rm turb}/|E_{\rm g}|$ & $0.03$\\
   \hline
  \end{tabular}
\end{table*}

Previous simulations, however, overlooked the influence of magnetic fields, which could potentially alter the characteristics of binary and multiple stellar systems. In the early Universe, theoretical investigations suggested the existence of minute seed magnetic fields generated from cosmological phenomena (e.g., \cite{Turner_Widrow1988}; \cite{Quashnock1989}; \cite{Ratra1992}; \cite{Wagstaff2014}; \cite{Saga2015}; \cite{Subramanian2016})
and astrophysical processes, such as the Biermann battery mechanism (\cite{Biermann1950}; \cite{Biermann1951})
associated with supernova (SN) explosions (e.g,, \cite{Hanayama2005}), radiation effects (e.g., \cite{Gnedin2000}; \cite{Langer2003}; \cite{Doi_Susa2011}; \cite{Attia2021}), galaxy formation (e.g., \cite{Kulsrud1997}), and the streaming of primordial cosmic rays (\cite{Ohira2020}, \yearcite{Ohira2021}). In particular, accretion shocks onto minihalos can generate weak magnetic fields in first star-forming regions, with field strengths ranging from $10^{-20}$ to $10^{-18}\ \rm G$ (e.g., \cite{Xu2008}; \cite{McKee2020}).

Seed magnetic fields can be amplified further via a combined effect of global compression and dynamo action (e.g., \cite{Brandenburg_Subramanian2005}; \cite{Federrath2016}) during gravitational collapse. While compression alone might not suffice to generate magnetic fields that significantly influence gas dynamics, the turbulence naturally arising during gas accretion onto minihalos (e.g., \cite{Greif2012}; \cite{Stacy_Bromm2013}; \cite{McKee2020}) can drive a small-scale dynamo, thereby enhancing the amplification process (e.g., \cite{Batchelor1950}; \cite{Kazantsev1968}; \cite{Kulsrud_Anderson1992}; \cite{Schekochihin2004}; \cite{Schleicher2010}; \cite{Sur2010}; \cite{Turk2012}; \cite{Schober2012}; \cite{Xu_Lazarian2016}; \cite{McKee2020}; \cite{Sadanari2023}; \cite{Higashi2024}). 
As a result, these fields can reach strengths comparable to turbulent energy levels before protostar formation (e.g., \cite{Schober2012}; \cite{McKee2020}; \cite{Higashi2024}). Hence, we can expect that the first star-forming clouds are magnetized close to the equipartition level, exhibiting a perturbed morphology as a result of turbulent motions, i.e., turbulent magnetic fields.

The impact of magnetic fields on star formation has been actively studied in the context of present-day star forming regions, 
where strong and coherent magnetic fields have been observed (e.g., \cite{Crutcher1999}; \cite{Troland_Crutcher2008}; \cite{Pillai2015}).
According to magnetohydrodynamics (MHD) simulations, these coherent magnetic fields can efficiently transport angular momentum
through processes such as the magnetic braking (e.g., \cite{Gillis1974}, \yearcite{Gillis1979}; \cite{Mouschovias_Paleologou1979}), MHD outflows (e.g., \cite{Blandford_Payne1982}; \cite{Tomisaka2002}; \cite{Banerjee_Pudritz2006}; \cite{Machida2008a}; \cite{Tomida2010}), and turbulence generation 
via magneto-rotational instability (MRI) (e.g., \cite{Balbus_Hawley1991}; \cite{Bai_Stone2013}). These MHD phenomena result in the 
reduction of the disk size (e.g., \cite{Tomisaka2000}; \cite{Price_Bate2007}; \cite{Machida2011}), suppression of fragmentation, decreased binary separation (e.g., \cite{Matsumoto2024}), and lower star formation efficiency (e.g., \cite{Padoan_Nordlund2011}; \cite{Federrath_Klessen2012}; \cite{Machida_Hosokawa2013}; \cite{Federrath2015}).

In turbulent magnetic fields, as expected in the first star-forming regions, 
numerical simulations suggest that the efficiency of magnetic braking tends to decrease due to the 
effects of the complex field structure (\cite{Hennebelle_Ciardi2009}; \cite{Seifried2012}; \cite{Joos2012}; \cite{Hirano2020}) 
and magnetic dissipation via turbulent reconnection (e.g., \cite{Santos-Lima2012}; \cite{Joos2013}).
Additionally, the power of MHD outflows is weaker in the presence of turbulent fields 
(e.g., \cite{Matsumoto2017}; \cite{Lewis_Bate2018}; \cite{Hirano2020}; \cite{Mignon2021b}; \cite{Takaishi2024}). 
\citet{Gerrard2019} also pointed out that without a coherent magnetic field, outflows may not occur.

Several authors have performed MHD simulations of first-star formation during the accretion phase
to investigate the impact of turbulent fields generated via dynamo on disk fragmentation and the 
initial mass function (IMF) (e.g., \cite{Sharda2020}, \yearcite{Sharda2021}; \cite{Stacy2022}; \cite{Prole2022}).
However, these simulations yield conflicting results. Simulations which consider the magnetic field amplification from 
the collapse phase indicate that disk fragmentation is suppressed, resulting in a top-heavy IMF 
(\cite{Sharda2020}, \yearcite{Sharda2021}; \cite{Stacy2022}). Conversely, \citet{Prole2022} observed no magnetic effects on the 
fragmentation process when introducing random fields dominated by small-scale power spectrum, as predicted 
by dynamo theory, at the end of collapse phase. The discrepancy may arise from different magnetic field structures at 
the beginning of the accretion phase, but this remains unclear.
 
Here, we perform the MHD simulations starting from the gravitational collapse of a turbulent gas cloud core, 
during which the magnetic fields are amplified via both dynamo and compression mechanisms. Subsequently, we 
simulate the ensuing evolution up to the accretion phase, where disk fragmentation occurs frequently, 
giving rise to the formation of multiple systems. Our investigation focuses on three key magnetic effects: 
magnetic pressure, previously demonstrated to stabilize the disk (e.g., \cite{Stacy2022}); magnetic torques, 
responsible for angular momentum transport; and MHD outflows, capable of ejecting both mass and angular 
momentum. We investigate how these effects influence disk size, disk fragmentation, and binary properties 
such as separation, aiming to identify the conditions necessary for magnetic effects to impact the formation 
of first stars.
 
This paper is organized as follows:
Section \ref{sec_method} outlines the numerical methods we used and the initial setup.
In section \ref{result}, we present the results of our simulations. We first explain the magnetic 
amplification during the collapse in section \ref{sec2.3}. Subsequently, in section \ref{secDisk_evo}, 
we examine the evolution of the disk and its fragmentation during the accretion phase. 
Then, we examine how magnetic fields evolved within the disk and explore their effects on 
the system, focusing on magnetic pressure, magnetic torques, and MHD outflow in section 
\ref{secBeffects}. In section \ref{secNfrag}, we discuss how these magnetic effects influenced 
the properties of binary and multiple systems. We summarize our findings and discuss their 
implications for the formation of first stars in section \ref{secSum}.
\begin{table*}
 \caption{Model parameters.}
 \label{table_model}
 \centering
  \begin{tabular}{lcccccc}
   \hline \hline
   Model  &  $ E_{\rm{rot}}/|E_{\rm g}|$  &  $E_{\rm{turb}}/|E_{\rm g}|$  &  $E_{\rm{mag}}/|E_{\rm g}|$ & $\mu_{0}$ & $B_{\rm init}\ \rm[G]$ & B-field @ protostar formation\\
   \hline
   T2M0   ..... & $10^{-2}$  & $3\times 10^{-2}$   &  $0                      $    &  $\infty$      &  $ 0         $    &   - \\
   T2M7   ..... & $10^{-2}$  & $3\times 10^{-2}$   &  $2\times 10^{-7}$    &  $27000$    &  $10^{-8}$     &  $B < B_{\rm eq}$  \\
   T2M5 .....   & $10^{-2}$  & $3\times 10^{-2}$   &  $2\times 10^{-5}$    &  $2700$      &  $10^{-7}$     &  $B\sim B_{\rm eq}$ ($n_{\rm H}\gtrsim10^{13}\ \rm cm^{-3}$)  \\
   T2M4 .....   & $10^{-2}$  & $3\times 10^{-2}$   &  $2\times 10^{-4}$    &   $100$       &  $5\times10^{-7}$     & $B\sim B_{\rm eq}$ ($n_{\rm H}\gtrsim10^{11}\ \rm cm^{-3}$)\\
   \hline
  \end{tabular}\\
   Note.$-$ The dimensionless parameter $\mu_{\rm 0}$ indicates the mass-to-flux ratio normalized by the critical value $(M/\Phi)_{\rm cr}$. 
\end{table*}
\section{Numerical Method}\label{sec_method}
\subsection{Code description}
We perform three-dimensional ideal MHD simulations using the adaptive mesh refinement 
(AMR) code SFUMATO (\cite{Matsumoto2007}; \cite{Matsumoto2015}). These simulations 
start with the collapse of a gas cloud core and extend though the accretion phase. To model 
the chemical and thermal evolution, we use the SFUMATO-RT module (\cite{Sugimura2020}).
The chemical network and the thermal process are described in detail in \cite{Sadanari2021}.
Since non-ideal MHD effects are expected to be negligible in primordial gas 
(e.g., \cite{Maki_Susa2004}, \yearcite{Maki_Susa2007}; \cite{Machida2008}; \cite{Sadanari2023}), 
we  adopt ideal MHD assumptions in our simulations. We solve the same governing equations as 
in \citet{Sadanari2021}, using the HLLD approximate Riemann solver (\cite{Miyoshi_Kusano2005}) for 
the MHD part with a divergence cleaning techniques (\cite{Dedner2002}) and the multi-grid method for 
the self-gravity part. Both parts are solved with a second-order accuracy in space and time. 

We initially set the box size and the number of cells in each direction to 
$L_{\rm box}=4\times10^{6}\ \rm au$ and $N_{\rm base}=256$, respectively.  We employ a cell 
refinement condition where the cells are divided if their size exceeds $1/64$ of the local Jeans length, 
to ensure accurate representation of dynamo amplification during the collapse phase 
(e.g., \cite{Sur2010}; \cite{Federrath2011b}; \cite{Turk2012}). The maximum refined level is limited 
to $l_{\rm max}=15$, and the minimum cell size is $\Delta x_{\rm min}\simeq4.7\times 10^{-1}\ \rm au$.

To simulate up to the accretion phase without employing the sink particle technique, we suspend cooling 
when the density reaches $n_{\rm th}=10^{14}\ \rm cm^{-3}$. This prevents further collapse and results in 
the formation of artificial adiabatic gas cores with $n>n_{\rm th}$. Although these cores are considerably 
larger than the radius of real protostars, we interpret their formation as indicative of protostar formation. 
Our simulations continue for at least $3500\ \rm yr$ after the formation of the primary protostar 
(adiabatic core) to observe the magnetic effects on disk fragmentation.
\subsection{Initial properties of gas cloud cores}
\label{sec_init}
As in our previous paper \citet{Sadanari2023}, we consider a single gas cloud core in the loitering phase, 
on the verge of gravitational collapse. We vary the magnetic field strength as a parameter for each 
simulation run. We adopt a density distribution characterized by a gravitationally unstable Bonnor-Ebert 
sphere (\cite{Ebert1955}; \cite{Bonnor1956}), with a density profile enhanced by a factor of $1.4$ to initiate 
collapse. 

This core is initially endowed with rigid rotation and turbulence, characterized by a typical power spectrum indicative of 
supersonic and compressible turbulence, given by $P(k)\propto k^{-4}$ (e.g., \cite{Federrath2021}), where $k$ denotes the 
wavenumber. The rotational and turbulent energies within the core amount to 1\% and 3\% of the gravitational 
energy $|E_{\rm grav}|$, respectively, consistent with typical energies anticipated from cosmological 
simulations (\cite{Greif2012}; \cite{Stacy_Bromm2013}; \cite{Stacy2022}). Note that the choice of initial 
turbulence energy does not significantly impact the simulation results because turbulence can be 
amplified during the gravitational collapse (e.g., \cite{Vazquez1998}; \cite{Federrath2011b}; \cite{Higashi2021}, 
\yearcite{Higashi2022}). The various quantities characterizing the cloud core employed as the initial 
condition are summarized in Table \ref{table_cloud}.

We introduce uniform magnetic fields aligned with the rotation axis, each possessing distinct levels of 
magnetic energy: $E_{\rm mag}/|E_{\rm grav}| = 2\times10^{-7}$, $2\times10^{-5}$, and $2\times10^{-4}$. 
Correspondingly, the initial strengths of these magnetic fields are set to $B_{\rm init} = 10^{-8}\ \rm G$, 
$10^{-7}\ \rm G$, and $5\times10^{-7}\ \rm G$, respectively. In addition, for comparison, we conduct 
hydrodynamics simulations devoid of any magnetic field. These initial field strengths are chosen to allow 
for the dynamo process during collapse, ensuring that the magnetic energy remains lower than the 
turbulent energy. By adjusting the initial field strengths, we replicate various magnetic field strengths 
and structures at the epoch of the end of the collapse phase. This enables us to investigate the necessary 
conditions under which turbulent magnetic fields impact the formation of first stars.
\begin{figure*}
\begin{center}
\includegraphics[width=160mm]{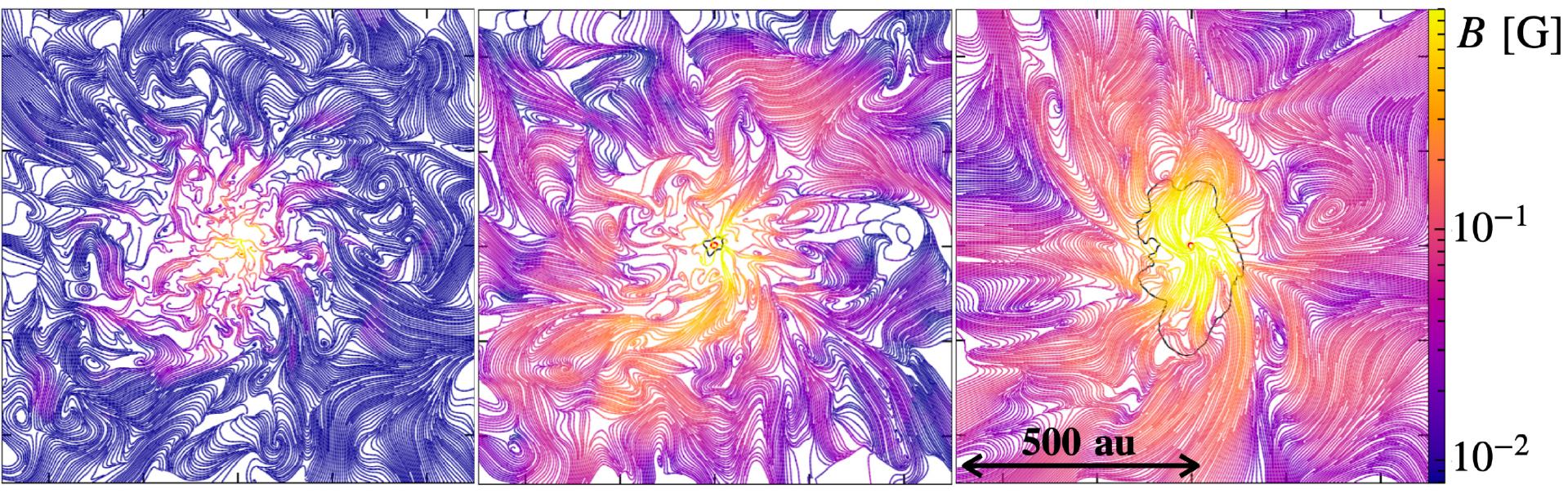}
\end{center} 
\caption{
Configuration of sliced magnetic field lines on the face-on plane at the epoch of the formation 
of the primary protostar for the cases of $B_{\rm init}=10^{-8}\ \rm G$ (left), $10^{-7}\ \rm G$ 
(middle), and $5\times10^{-7}\ \rm G$ (right). The field lines are the 2D projection
of the 3D magnetic fields onto the plane. The protostar is located at the center of each panel. 
Line color indicates the field strength, and the areas enclosed by black lines in the middle 
and right panels indicate the region where the fields reach a full equipartition state.
}
\label{fig4.-1}
\end{figure*}
\section{Results}\label{result}
\subsection{Magnetic amplification during the collapse phase}\label{sec2.3}
Firstly, we investigate the evolution of magnetic fields during the collapse 
phase and analyze the field structure at the onset of the accretion phase in our simulations. 
According to theoretical prediction, when the initial magnetic energy is lower than the turbulent 
energy, a small-scale dynamo driven by turbulence can enhance the magnetic amplification rate.

The dynamo process during the collapse phase can be divided into three 
distinct stages (e.g., \cite{Xu_Lazarian2016}; \cite{McKee2020}). Initially, during the kinematic stage, magnetic fields 
undergo exponential amplification on small scales, without encountering significant back reaction from 
magnetic forces. This stage is characterized by the prevalence of small-scale structures in the magnetic 
field configuration. Subsequently, in the nonlinear stage following equipartition on the smallest 
scales, further dynamo amplification on small scales is impeded by the back reaction. However, 
on larger scales, magnetic fields continue to amplify through dynamo action, eventually becoming 
dominant over the small-scale fields. 
In the final phase, once all scales up to the driving scale, corresponding to the Jeans scale, 
the dynamo amplification halts.
At this stage, the field strength approaches to the equipartition with turbulence, defined as
\begin{eqnarray}
\label{eq_Beq}
B_{\rm eq} \equiv \sqrt{4\pi \rho} V_{\rm turb}, 
\end{eqnarray}
where $V_{\rm turb}$ denotes turbulent velocity. Note that the saturation of the turbulent dynamo depends on the properties of turbulence, i.e., the Mach number and the driving mode (e.g., \cite{Federrath2011a}) and on the plasma parameter, i.e., Prandtl number 
 (\cite{Federrath2014}). For subsonic turbulence, the magnetic fields typically saturate around  $0.7B_{\rm eq}$ 
(\cite{Haugen2004}; \cite{Federrath2011a}; \cite{Brandenburg2014}).
During this phase, the magnetic energy becomes 
concentrated on the largest scales (Jeans scale) as the magnetic power spectrum follows the 
turbulent spectrum $P(k) \propto k^{-4}$. Additionally, \citet{Sadanari2023} observed a tendency 
for field lines to align due to the effects of global compression of equipartition fields.

Figure \ref{fig4.-1} shows the configuration and strength of sliced magnetic field lines on the plane perpendicular 
to the rotation axis at the moment of first (primary) protostar formation for three magnetized cases. 
The field lines are 2D projection of the 3D magnetic fields onto the plane.
These figures illustrate the generation of turbulent magnetic fields in all cases, attributable to dynamo action. We can also see 
that the surrounding field strength is greater in cases where the initial magnetization is higher. 

In the least magnetized case, with $B_{\rm init} = 10^{-8}\ \rm G$, magnetic fields are amplified until 
their energy matches that of the turbulent energy on the smallest resolved scale, i.e., the cell scale, by this moment.
At this stage, as magnetic fields on larger scales have not yet reached equipartition with the turbulent energy, 
the predominant field energy remains concentrated on smaller scales. Consequently, only turbulent motion on 
smaller scales is affected by the back reaction of magnetic forces.

In the cases with stronger initial magnetic fields, $B_{\rm init} = 10^{-7}\ \rm G$ and $5\times 10^{-7}\ \rm G$, the magnetic fields achieve equipartition levels before the formation of the protostar. Consequently, the magnetic energy becomes concentrated on larger scales, specifically the Jeans scale. The density at which the magnetic fields reach equipartition is approximately $10^{13}\ \rm cm^{-3}$ for $B_{\rm init} = 10^{-7}\ \rm G$ and $10^{11}\ \rm cm^{-3}$ for $B_{\rm init} = 5\times10^{-7}\ \rm G$, respectively. Thus, the spatial extent where $B\sim B_{\rm eq}$ is greater for the case of $B_{\rm init} = 5\times10^{-7}\ \rm G$. This region is illustrated in figure \ref{fig4.-1} by the enclosed area marked by black thin lines. The properties of each magnetized case are summarized in Table \ref{table_model} 

\begin{figure*}
\begin{center}
\includegraphics[width=160mm]{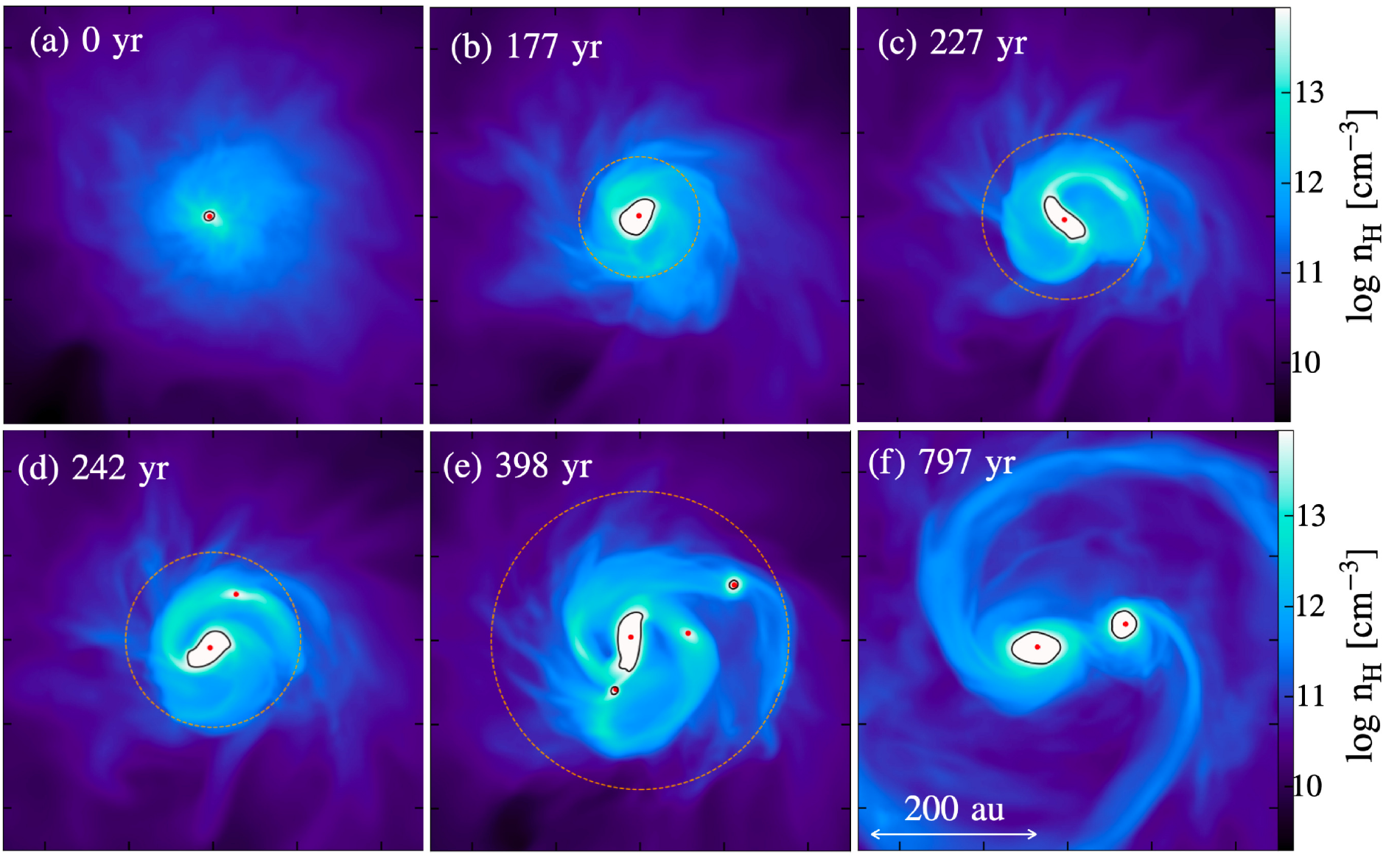} 
\end{center}
\caption{
Sequence of face-on density-weighted projections illustrating the distribution of number 
density at six different time intervals, from the formation of a primary protostar (panel a) 
to the emergence of a binary system (panel f) in the unmagnetized case ($B_{\rm init}=0\ \rm G$). 
The time elapsed from the formation of the primary protostar is indicated in the top left corner of 
each panel. The regions enclosed by the black lines and the red points delineate the adiabatic regions, 
where $n_{\rm H}>10^{14}\ \rm cm^{-3}$ and their center of mass, respectively. 
The orange circles indicate the disk region, defined as the transition radius $R$ where the azimuthally-averaged rotation speed $v_{\phi}(R)$ 
exceeds three times the collapse speed $v_{R}(R)$. we do not draw the orange circle in the panel (f)
because the $R_{\rm disk}$ is larger than one side of the snapshot.
}
\label{fig4.0}
\end{figure*}
\begin{figure*}
\begin{center}
\includegraphics[width=160mm]{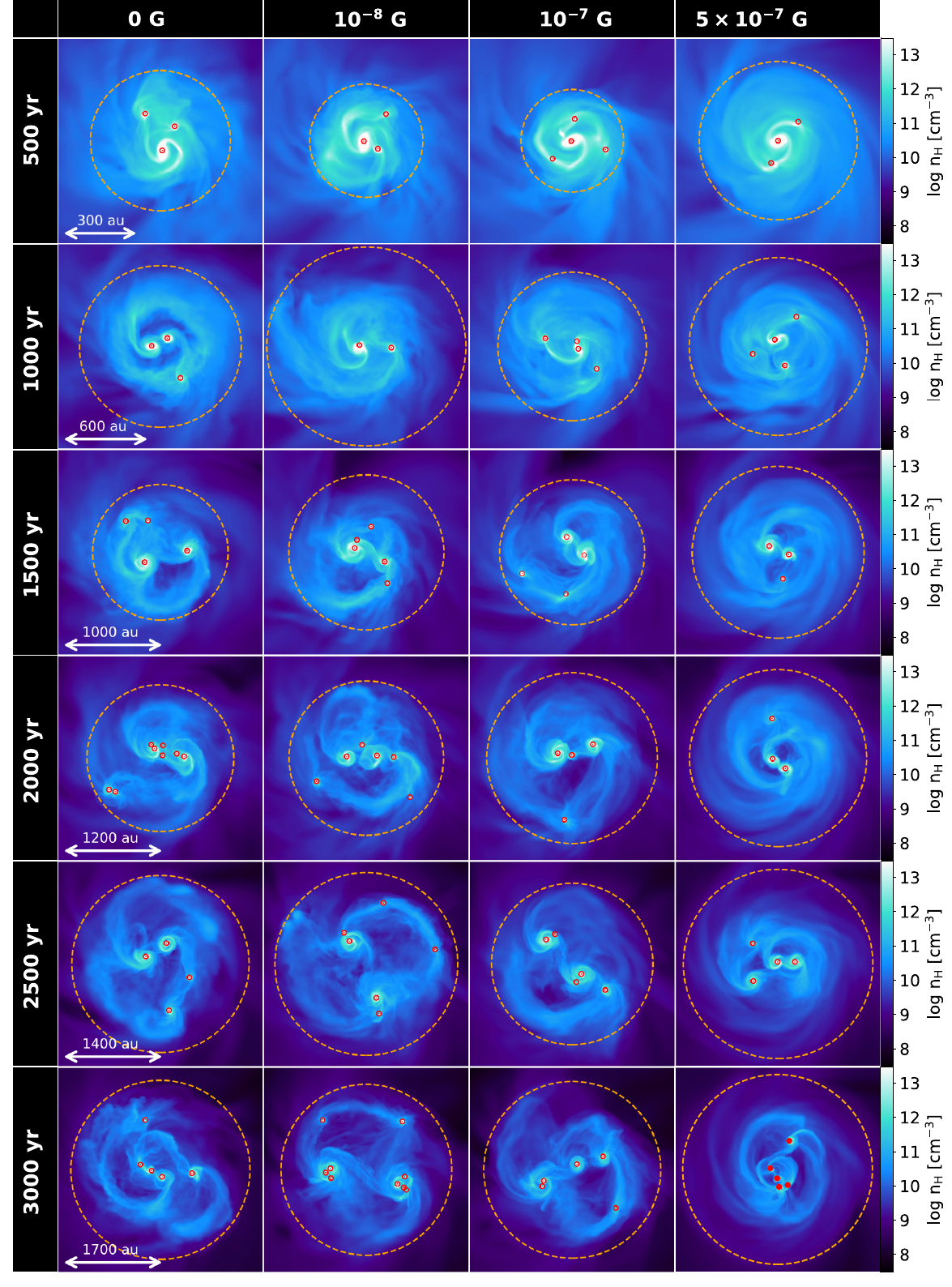} 
\end{center}
\caption{
The face-on mass-weighted average density distribution at six different time intervals:
$\rm 500\ yr,\ 1000\ yr,\ 1500\ yr,\ 2000\ yr,\ 2500\ yr,$ and $\rm 3000\ yr$ after the 
formation of the primary protostar, with four different initial magnetic field strengths: 
$B_{\rm init}=0\ \rm G,\ 10^{-8}\ \rm G,\ 10^{-7}\ G,$ and $5\times10^{-7}\ \rm G$ (from 
left to right columns). The center of each snapshot corresponds to the center of mass. 
Snapshots at the same epoch (row) share the same scale, as indicated by the white arrow 
in the left panel. The orange circles indicate the radius of the disk region $R_{\rm disk}$.
}
\label{fig4.1}
\end{figure*}
\begin{figure*}
\begin{center}
\includegraphics[width=160mm]{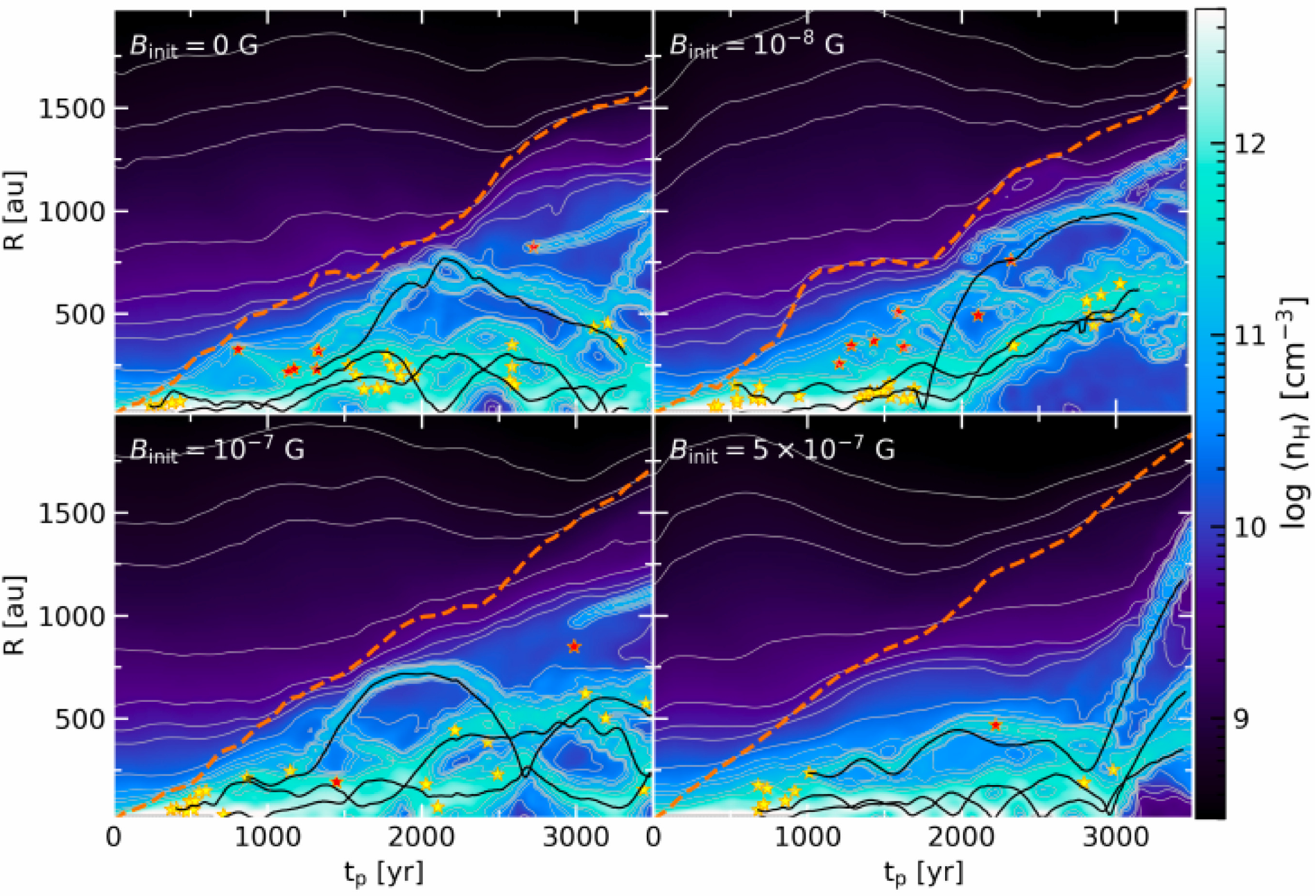} 
\end{center}
\caption{
The time evolution of the radial profile of azimuthally-averaged density for four different cases: $B_{\rm init}=0\ \rm G,\ 10^{-8}\ \rm G,\ 10^{-7}\ G,$ and $5\times10^{-7}\ \rm G$. Orange dashed lines represent the disk radius $R_{\rm disk}$, while black lines indicate the trajectories of three massive protostars. 
Yellow (red) stars indicate the radial positions at which protostars form through the fragmentation of spiral arms in circumstellar (circum-multiple/binary, respectively) disks. 
Thin lines represent the contour of the azimuthally-averaged density.
}
\label{fig4.4}
\end{figure*}
\subsection{Disk evolution and fragmentation}\label{secDisk_evo}
Next, we investigate the gas dynamics during the accretion phase following the formation of the primary protostar. 
First, we focus on examining how circumstellar disks around primary protostars fragment to give 
rise to binary or multiple systems in the absence of an initial magnetic field ($B_{\rm init}=0\ \rm G$). 
As we will see later, the fragmentation process of the disks proceeds similarly even in the magnetized cases.

Figure \ref{fig4.0} presents a sequence of face-on density-weighted projections showing the evolution from the formation 
of a primary protostar to the emergence of a binary system. The enclosed area marked by black lines corresponds to the 
adiabatic core or protostar ($n_{\rm H}>10^{14}\ \rm cm^{-3}$), with its center of gravity indicated by a red dot. 
Additionally, we show the disk region where rotational motion dominates over collapse motion with orange circles in 
each panel. These circles are determined based on the disk radius $R_{\rm disk}$, defined as the transition radius 
$R$ where the azimuthally-averaged rotation speed $v_{\phi}(R)$ surpasses three times the collapse speed $v_{R}(R)$. 
This empirical criterion effectively traces the disk region.

Initially, a primary protostar emerges at the center of the collapsing gas cloud, as illustrated in figure \ref{fig4.0}(a). 
Subsequently, it gains mass through the accretion of envelope gas. The envelope accretes onto the center along with 
the angular momentum inherited from the initial rotation and turbulence of the cloud core, leading to the spontaneous 
formation of a rotating disk around the protostar, as depicted by the orange circle in figure \ref{fig4.0}(b).

As time elapses, the disk region continues to expand and gain mass, eventually becoming gravitationally unstable. This 
instability manifests in the emergence of two spiral arms within the disk, as shown in figure \ref{fig4.0}(c). Quantitatively, 
the disk's gravitational instability can be assessed using the Toomre $Q$ parameter (\cite{Toomre1964}), 
defined as $Q \equiv c_{\rm s}\Omega_{\rm epi}/(\pi G \Sigma)$, where $c_{\rm s}$, $\Omega_{\rm epi}$, and $\Sigma$ are
 the sound speed, epicyclic frequency, and surface density of the disk, respectively.
When the Toomre $Q$ parameter approaches unity, spiral arms begin to emerge within the disk 
(e.g., \cite{Laughlin_Bodenheimer1994}), a criterion we also confirm in our simulations. These arms 
play a crucial role in redistributing mass and angular momentum within the disk by generating non-axisymmetric 
gravitational torques, thereby aiding in disk stabilization.

However, in cases characterized by high accretion rates such as in our case of first star formation, gravitational 
instability persists and intensifies as the disk accumulates mass more rapidly than it can be stabilized by the spiral 
arms. Consequently, the spiral arms continue to acquire mass, eventually becoming gravitationally unstable and 
undergoing fragmentation, as evidenced by the red point in figure \ref{fig4.0}(d). As noted by \citet{Takahashi2016}, spiral 
arms become unstable when the Toomre $Q$ parameter drops below approximately 0.6, as inferred from linear stability 
analyses. In most cases, this condition is met when the spiral arms are tightly wound and collide with each other, 
causing an increase in the surface density. Consequently, the spiral arms continue fragmenting further, resulting in the 
emergence of three protostars (fragments) surrounding the primary protostar, as depicted in figure \ref{fig4.0}(e). 
However, gravitational torques exerted by the spiral arms cause most of the protostars to migrate towards the primary 
protostar. Eventually, only one companion protostar survives around the central protostar, resulting in the formation of
a binary system, as illustrated in figure \ref{fig4.0}(f).

Following the formation of binary or multiple systems, we can identify two distinct types of disks: circumstellar disks surrounding each individual protostar, and circum-binary or circum-multiple disks enveloping the entire binary or multiple systems. The circum-multiple disk primarily gains mass from the accreting envelope, whereas the circumstellar disk primarily receives gas from the spiral arms extending from each protostar, as depicted in figure \ref{fig4.0}(f).
As time elapses, the spiral arms in both types of disks will undergo further fragmentation, as detailed later.

Next, we also examine the magnetized cases. Figure \ref{fig4.1} presents a face-on view of the density structure for cases with $B_{\rm init}=0\ \rm G,\ 10^{-8}\ \rm G,\ 10^{-7}\ G,$ and $5\times10^{-7}\ \rm G$, spanning from $\rm 500\ yr$ to $\rm 3000\ yr$ at intervals of $\rm 500\ yr$. Additionally, the time evolution of the radial density profile for all these cases is depicted in figure \ref{fig4.4}. The horizontal axis represents the time after the formation of the primary protostar $t_{\rm p}$, while the vertical axis represents the radius from the center of mass $R$. The color scale represents the azimuthally-averaged density on the disk midplane. In figures. \ref{fig4.1} and \ref{fig4.4}, orange circles and orange dashed lines respectively denote the disk region, defined in the same manner as in figure \ref{fig4.0}.
Additionally, in figure \ref{fig4.4}, yellow and red star symbols indicate the positions where fragmentation occurred in the circumstellar and circum-multiple disks, respectively.

From figure \ref{fig4.1}, we can clearly see that the binary or multiple systems 
form in all cases, regardless of magnetized level. These systems form through 
the fragmentation process of the spiral arms, as seen in the unmagnetized case.
Once multiple systems are established, fragmentation primarily occurs in two 
distinct regions: one involves the fragmentation of the spiral arms within the 
circumstellar disk (yellow stars in figure \ref{fig4.4}), while the other 
involves the fragmentation of the spiral arms extending outward from the binary 
systems (red stars in figure \ref{fig4.4}).

The former type of fragmentation primarily occurs across the all cases, 
albeit with a decreased frequency in more strongly magnetized cases, 
as evidenced by the diminishing number of yellow stars in figure \ref{fig4.4}. 
This reduction is mainly attributed to the stabilizing effect induced by 
magnetic pressure (section \ref{secPmag}). Protostars formed through this fragmentation process
often merge with the central protostar due to their initial proximity. 

In the latter type of fragmentation, fragmentation predominantly occurs 
in the relatively outer region of the disk (as indicated by the red 
stars in figure \ref{fig4.4}). In the weakly magnetized cases 
($B_{\rm init}=0\ \rm G$ and $10^{-8}\ \rm G$), the spiral arms extend 
outward to the outer region, facilitating fragmentation. However, in the 
case of $B_{\rm init}=10^{-7}\ \rm G$, although prominent spiral arms are 
observed in the circum-multiple disk, fragmentation frequency is reduced 
by magnetic pressure effect compared to the weakly magnetized cases, as 
evidenced by the fewer red stars in figure \ref{fig4.4}. In the strongest 
magnetized case ($B_{\rm init}=5\times10^{-7}\ \rm G$), prominent spirals 
are absent due to the angular momentum transport by magnetic torques (see 
section \ref{secBtorques}), resulting in the suppression of fragmentation 
in the outer region. Consequently, both the gas and protostars tend to be 
more concentrated toward the center of the disk region compared to other 
weakly magnetized cases ($B_{\rm init} \leq 10^{-7} \ \rm G$), as illustrated 
in figures. \ref{fig4.1} and \ref{fig4.4}.

\subsection{Magnetic field evolution and their impact on the disks}
\label{secBeffects}
\begin{figure*}
\begin{center}
\includegraphics[width=160mm]{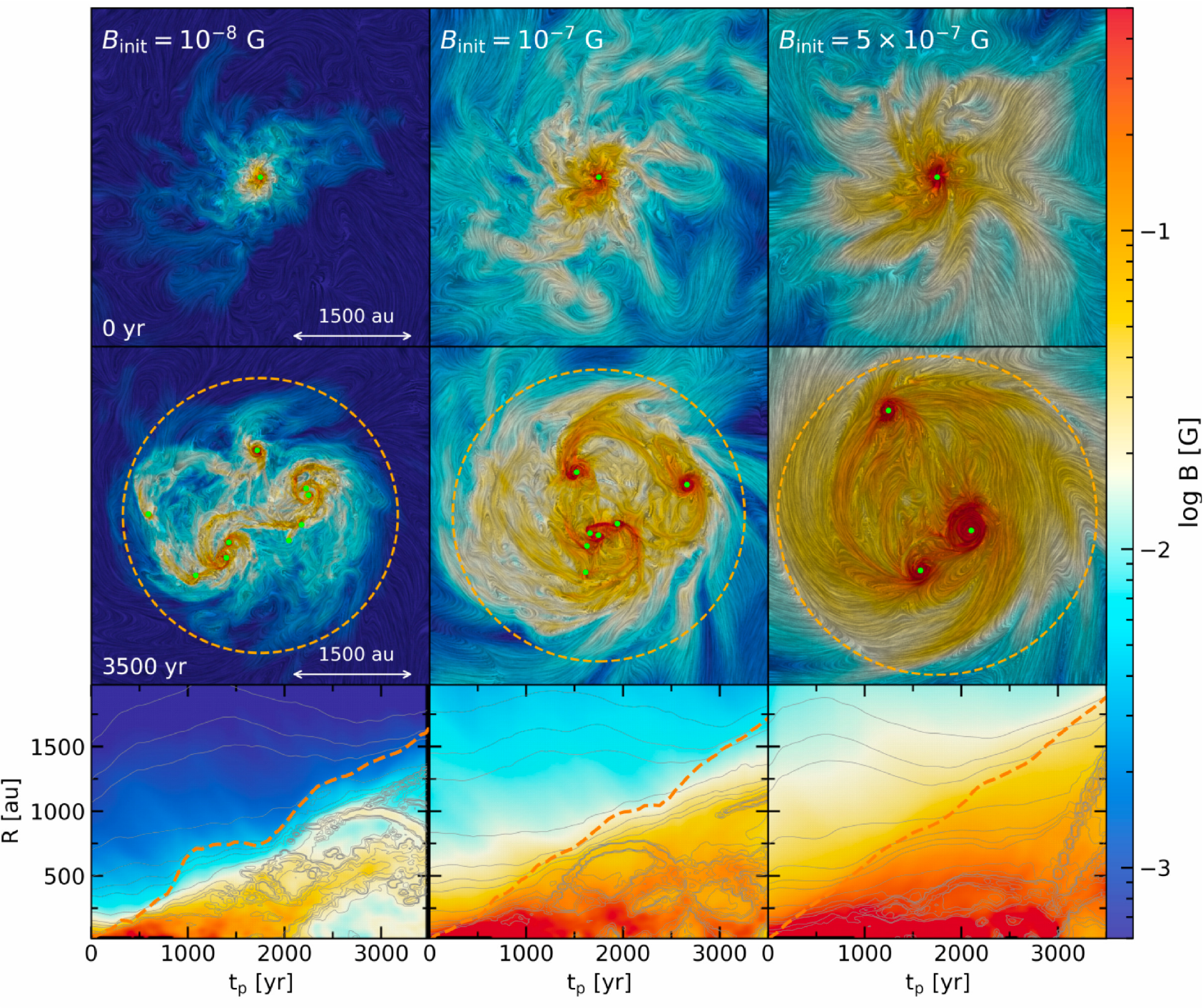} 
\end{center}
\caption{
The face-on, mass-averaged magnetic field strength distribution at $t_{\rm p}= 0\ \rm yr$ (top) and $3500\ \rm yr$ (middle) for the three magnetized cases: $B_{\rm init}=10^{-8}\ \rm G$ (left column), $10^{-7}\ \rm G$ (middle column), and $5\times10^{-7}\ \rm G$ (right column). The orange circles represent the radius of the disk region defined as in section \ref{secDisk_evo}. The positions of protostars in top and middle panels are indicated as green points. The shadow pattern represents the direction of magnetic field lines sliced in the plane. 
The bottom figures display the time evolution of the radial profile of azimuthally-averaged field strength, similar to figure \ref{fig4.4}. The orange dashed lines represent the radius of the disk region $R_{\rm disk}$. Thin gray lines in the bottom panels represent the density contour, as shown in figure \ref{fig4.4}.}
\label{fig4.6}
\end{figure*}
\begin{figure*}
\begin{center}
\includegraphics[width=160mm]{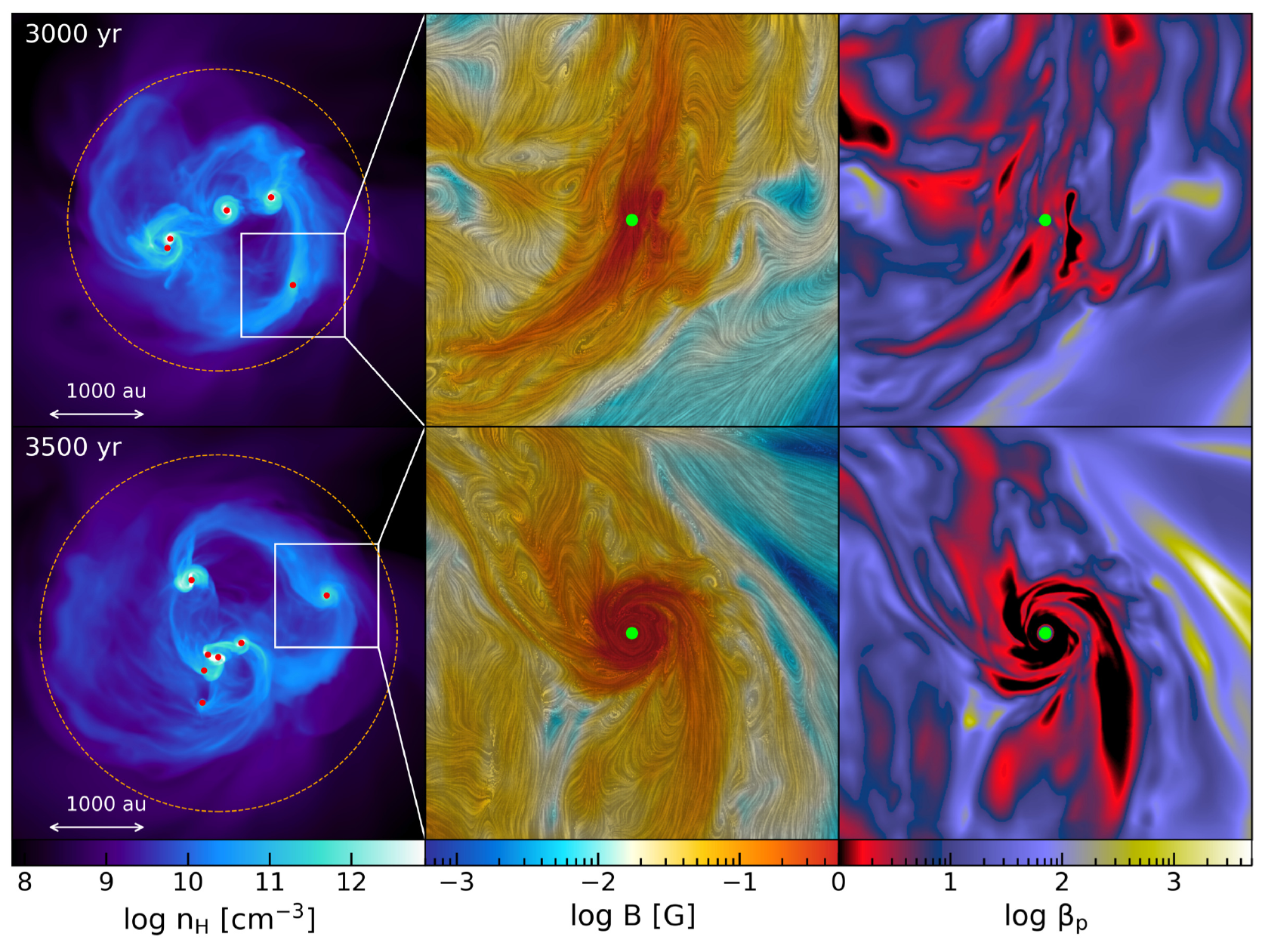} 
\end{center}
\caption{
Close-up view of the magnetic field structure around a protostar in the case of $B_{\rm init}=10^{-7}\ \rm G$. 
The left panel displays the density distribution of the entire disk region, with the boxed area highlighted for 
enlargement in the middle and right panels. The middle panel provides a close-up depiction of the magnetic 
field strength, while the right panel illustrates the plasma beta $\beta_{\rm p}\equiv P_{\rm th}/P_{\rm mag}$, 
specifically focusing on a spiral arm within the circum-multiple disk. The upper panels correspond to the 
moment when the arm fragments into a new protostar ($t_{\rm p}=3000\ \rm yr$), while the lower panels 
represent a later moment, $500\ \rm yr$ after the fragmentation. Shadow patterns in the field strength panel 
indicate the direction of field lines, as illustrated in figure \ref{fig4.6}.
}
\label{fig21Bmag}
\end{figure*}
\begin{figure}
\begin{center}
\includegraphics[width=80mm]{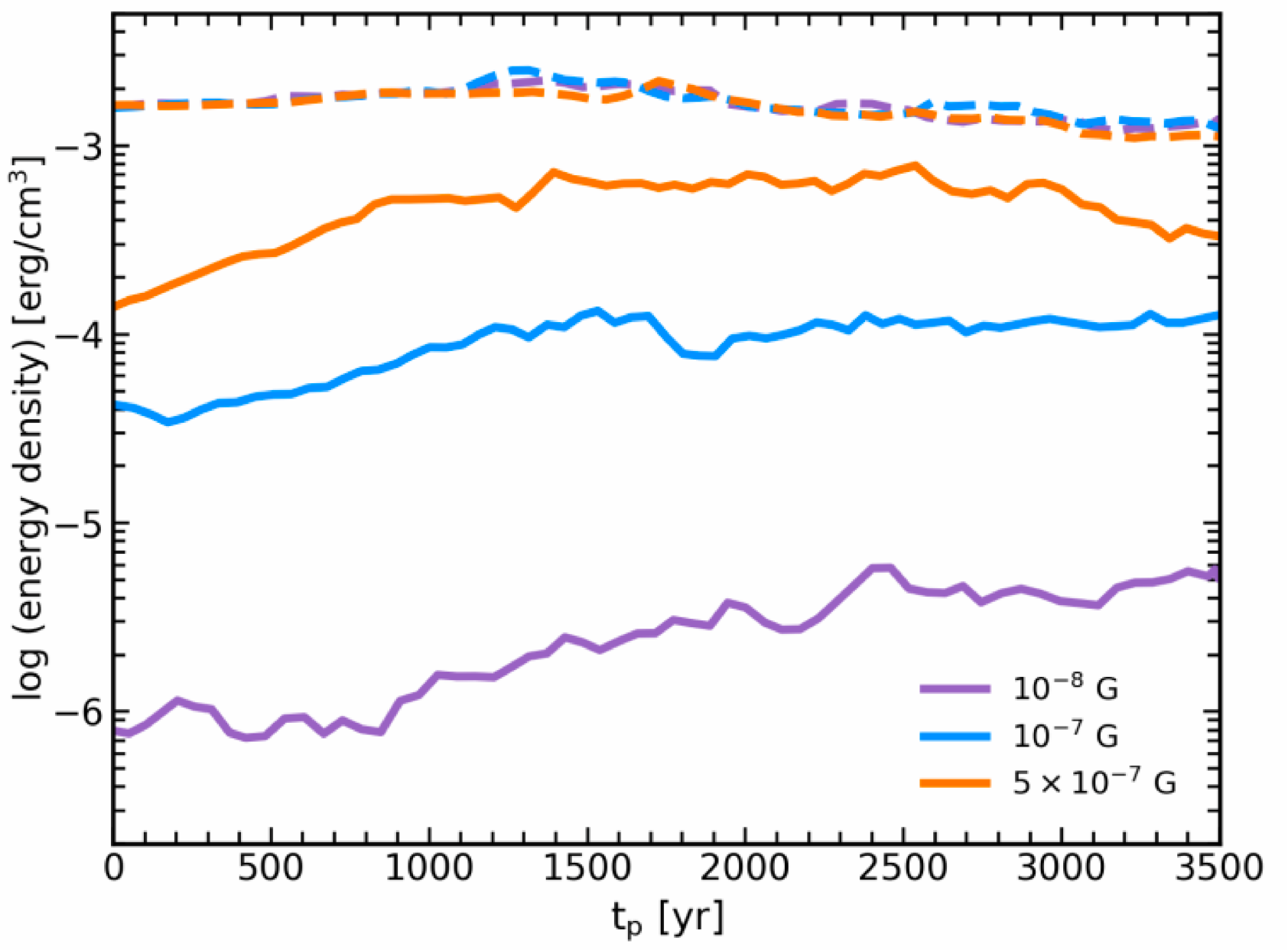} 
\end{center}
\caption{
Time evolution following the formation of the primary protostar ($t_{\rm p}$) 
of the average magnetic energy density within a specific density range 
from $6\times10^{9}\ \rm cm^{-3}$ to $8\times10^{9}\ \rm cm^{-3}$ (solid line). 
The thermal energy density is represented by dashed lines.
}
\label{fig4.5}
\end{figure}
\begin{figure*}
\begin{center}
\includegraphics[width=160mm]{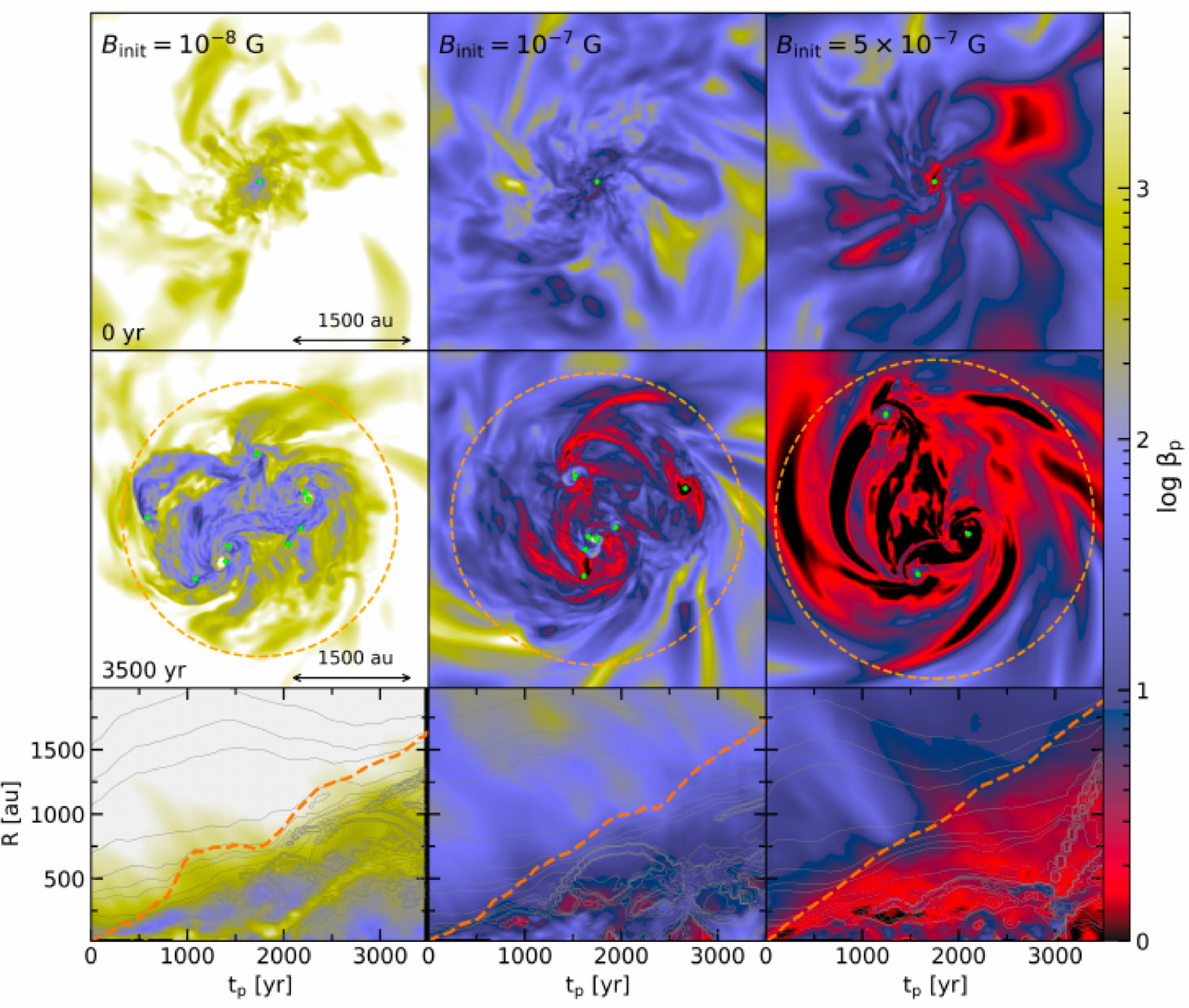} 
\end{center}
\caption{
The same as figure \ref{fig4.6}, but for the plasma beta $\beta_{\rm p}\equiv P_{\rm th}/P_{\rm mag}$.
}
\label{fig4.8}
\end{figure*}
Although the formation of multiple systems occurs regardless of the initial magnetic field strength, differences arise 
in the evolution and structure of the disk, leading to variations in the frequency of fragmentation, as observed in the 
previous section. These differences stem from the influence of magnetic fields on the gas dynamics within the disk 
region. In this section, we first investigate the evolution of magnetic field strength and configuration within the disk 
region. Subsequently, we examine three magnetic effects: magnetic pressure, magnetic torques as a mechanism for 
angular momentum transport, and magnetically-driven outflows, in this order.

\subsubsection{Magnetic evolution within the disk}\label{secBevo}
Figure \ref{fig4.6} shows the face-on density-weighted projection of magnetic field strength at $t_{\rm p}=0$ (top) and 
$t_{\rm p}=3500\ \rm yr$ (middle). The left column represents the case of $B_{\rm init}=10^{-8}\ \rm G$, the middle 
column $B_{\rm init}=10^{-7}\ \rm G$, and the right column $B_{\rm init}=5\times10^{-7}\ \rm G$. The orange circles 
and green points, respectively, indicate the disk region and the positions of protostars. The pattern of shadows 
in each snapshot represents the direction of the magnetic field lines sliced in the plane. 

As mentioned in section \ref{sec2.3}, the distribution of field strength at $t_{\rm p}=0$ (top panels in figure \ref{fig4.6}) 
varies depending on the initial field strength $B_{\rm init}$. A stronger initial field results in a higher field strength at the 
beginning of the accretion phase. Moreover, the magnetic field lines have undergone significant disturbance due to the 
dynamo amplification process during the collapse (also see figure \ref{fig4.-1}).

After the formation of the primary protostar, the rotational motion inside the disk region generates toroidal magnetic 
fields from poloidal fields, as illustrated in the shadow pattern in the middle panel in Fig \ref{fig4.6}. Particularly, the
differential rotation can amplify the toroidal fields, a phenomenon known as the $\Omega$ effect (e.g., \cite{Babcock1961}).  

Additionally, as the spiral arms extend outward, they gather toroidal fields along their trajectories, leading to 
the emergence of coherent toroidal fields within the arms. This is clearly seen in the middle top panel of 
figure \ref{fig21Bmag}, where a detailed view of the magnetic field structure around the spiral arm is depicted 
during its fragmentation into a new protostar for the magnetized case with $B_{\rm init}=10^{-7}\ \rm G$.
In this field configuration, the field strength within the arm experiences amplification due 
to compression along them, following a relation of $B\propto \rho$ at maximum. This amplification rate 
surpasses that in the case of spherical compression ($B\propto \rho^{2/3}$). Consequently, the average 
field strength within the disk is enhanced by compression along the arms after their emergence.

Another possible mechanism for magnetic amplification is the $\alpha$ dynamo process 
(e.g. \cite{Steenbeck1966}; \cite{Brandenburg_Subramanian2005}). This process involves 
the generation of poloidal fields from toroidal fields due to helical motion inside the disk. Some studies 
(e.g., \cite{Liao2021}) have argued that the $\alpha$ dynamo significantly contributes to field amplification 
in accretion disks, a conclusion supported by MHD simulations performed by \citet{Sharda2021}.
However, our simulations do not observe the growth of poloidal fields through this dynamo process, even 
in the outer region of the disk, where the required resolution is sufficient to capture the dynamo process 
(e.g., \cite{Federrath2011b}). We infer that disk fragmentation disrupts the helical motion, thereby impeding 
the driving mechanism of the $\alpha$ dynamo. Confirmation of this hypothesis will necessitate higher 
resolution in future numerical studies.

From the distribution of magnetic fields at $t_{\rm p}=3500\ \rm yr$ (middle panels of figure \ref{fig4.6}), 
we can find an increase in field strength within the disk region across the all magnetized cases. 
Comparing the magnetic fields distribution with the density (figure \ref{fig4.1}), we observe a rough 
correlation between them. In the bottom panels of figure \ref{fig4.6}, we present the time evolution of the 
radial profile of field strength, overlaid with the density contour from figure \ref{fig4.4} shown as thin gray lines. 
It appears that the field strength remains almost constant along the iso-density contour. 
This suggests that the amplification mechanisms shown above, i.e., $\Omega$ effect and $\alpha$ dynamo, 
are not efficient within the disk. 

While the magnetic field strength correlates well with the density, we also observe 
a slow amplification of the magnetic fields at a fixed density. To qualitatively analyze the 
magnetic evolution, we plot the time evolution of the average energy density of the magnetic 
field over the density range $6 \times 10^9 \ \rm cm^{-3}$ to $8 \times 10^9 \ \rm cm^{-3}$ 
for three magnetized cases in figure \ref{fig4.5}. Note that in this density regime, the field strength in 
none of these cases has reached the equipartition field $B_{\rm eq}$ by the beginning of the accretion 
phase. For the case with $B_{\rm init}=5\times10^{-7}\ \rm G$ (orange line), the magnetic field gradually 
amplifies initially and eventually reaches a level close to saturation, roughly equaling the thermal energy 
(indicated by the dashed line) within a factor of about two. For less magnetized cases 
($B_{\rm init}\leq 10^{-7}\ \rm G$), the field energies continue to increase slowly over time. This gradual 
amplification is caused by the combination of the $\Omega$ effect and the compression by spiral arms, 
as mentioned before. 

We speculate that the diffusion process of magnetic turbulent reconnection, 
discussed in previous studies (e.g., \cite{Santos-Lima2012}), slows down the amplification rates within 
the disk region. Additionally, disk fragmentation disturbs the magnetic field lines, further enhancing this 
diffusion mechanism (see figure \ref{fig4.6}). Consequently, the level of magnetization within the disk 
region during the early accretion phase, where the disk fragmentation frequently occurs, 
appears to depend on how strongly the magnetic fields are amplified during the collapse phase.
\subsubsection{Effects of magnetic pressure}\label{secPmag}
Magnetic pressure can stabilize the disks against gravitational instability by adding to thermal pressure. 
To investigate its impact on gas dynamics, we evaluate the plasma beta $\beta_{\rm p}$, which represents the 
ratio of thermal pressure ($P_{\rm th}$) to magnetic pressure ($P_{\rm mag}=B^{2}/(8\pi)$). 
As $\beta_{\rm p}$ approaches unity, magnetic pressure can affect the gas dynamics.
Since transonic turbulence is realized in our simulations, $\beta_{\rm p}$ can be approximated 
as $\beta_{\rm p} \sim (B_{\rm eq}/B)^{2}$, where $B_{\rm eq}$ denotes the fields strength at the equipartition 
with turbulent energy as defined in equation (\ref{eq_Beq}). Figure \ref{fig4.8} shows the distribution of 
$\beta_{\rm p}$ at $t_{\rm p}=0\ \rm yr$ (top)  and $3500\ \rm yr$ (middle), similar to figure \ref{fig4.6}, along 
with the time evolution of radial profile (bottom). Roughly speaking, the regions colored from red to black 
are thought to be where magnetic pressure is likely to influence the gas dynamics.

Given that the thermal structure remains independent of magnetization levels, differences in $\beta_{\rm p}$ reflect 
variations in field strength. In the case of $B_{\rm init} = 10^{-8}\ \rm G$, where the magnetic fields fail 
to reach $B_{\rm eq}$ during the collapse, the central $\beta_{\rm p}$ at $t_{\rm p}=0$ (top left panel in figure \ref{fig4.8}) 
drops to a maximum of $\beta_{\rm p} \sim 10^2$. This suggests that the magnetic pressure is insignificantly weak 
compared to the thermal pressure at the onset of the accretion phase. For the cases where the fields are amplified 
to $B_{\rm eq}$ such as $B_{\rm init} = 10^{-7}\ \rm G$ and $5\times10^{-7}\ \rm G$, central $\beta_{\rm p}$ 
approaches the magnetization level of $\beta_{\rm p} \sim (B_{\rm eq}/B)^2\sim 1$ in both cases (top middle and top right panels). Notably, in the case of $B_{\rm init} = 5\times10^{-7}\ \rm G$, the region where the magnetic pressure 
can affect the gas dynamics (colored from red to black) is broader due to the earlier amplification to $B_{\rm eq}$.

The magnetization level inside the disk region appears to remain relatively constant over time, as seen in the 
bottom panels of figure \ref{fig4.8}. Here, the azimuthally averaged $\beta_{\rm p}$ below the orange dashed 
line persist at approximately  $\beta_{\rm p}\sim 10^{2}-10^{3}$ ($B_{\rm init}=10^{-8}\ \rm G$), 
$\beta_{\rm p} \sim 10-10^{2}$ ($B_{\rm init}=10^{-7}\ \rm G$), and $\beta_{\rm p} \sim 1-10$ 
($B_{\rm init}=5\times10^{-7}\ \rm G$), respectively. This constancy arises because magnetic amplification 
within the disk is not notably efficient, as show in figure \ref{fig4.5}. Therefore, whether magnetic pressure 
can stabilize the disk during the earlier accretion phase depends on the magnetization level during the collapse 
phase. From the middle and top panels of figure \ref{fig4.8}, we can observe that only in the most magnetized 
case of $B_{\rm init}=5\times10^{-7}\ \rm G$, magnetic pressure can stabilize the entire disk region.

However, when we focus on the magnetization levels within spiral arms, we notice a decrease in 
$\beta_{\rm p}$. For instance, the third column of figure \ref{fig21Bmag} shows the distribution of 
$\beta_{\rm p}$ around a single spiral arm in the case of $B_{\rm init}=10^{-7}\ \rm G$. Initially, 
at the moment of a new protostar formation through fragmentation of the spiral arm, there is a clear 
decline in $\beta_{\rm p}$ along the arm (top right panel) attributed to the accumulation of toroidal 
fields generated by disk rotation (top middle panel). $500\ \rm yr$ later, the coherent fields along the 
arm are effectively amplified by the rotation of a circumstellar disk (bottom middle panel), with minimal 
dissipation through magnetic reconnection. Consequently, $\beta_{\rm p}$ within the circumstellar disk 
decreases further, dipping below 1 (black region in bottom right panel), indicating the dominance of 
magnetic pressure, which stabilize the disk and inhibits further fragmentation. This suggests that 
there is a significant influence of magnetic pressure on the evolution of circumstellar disks and protostars 
originating from the fragmentation of spiral arms. 
\begin{figure}
\begin{center}
\includegraphics[width=80mm]{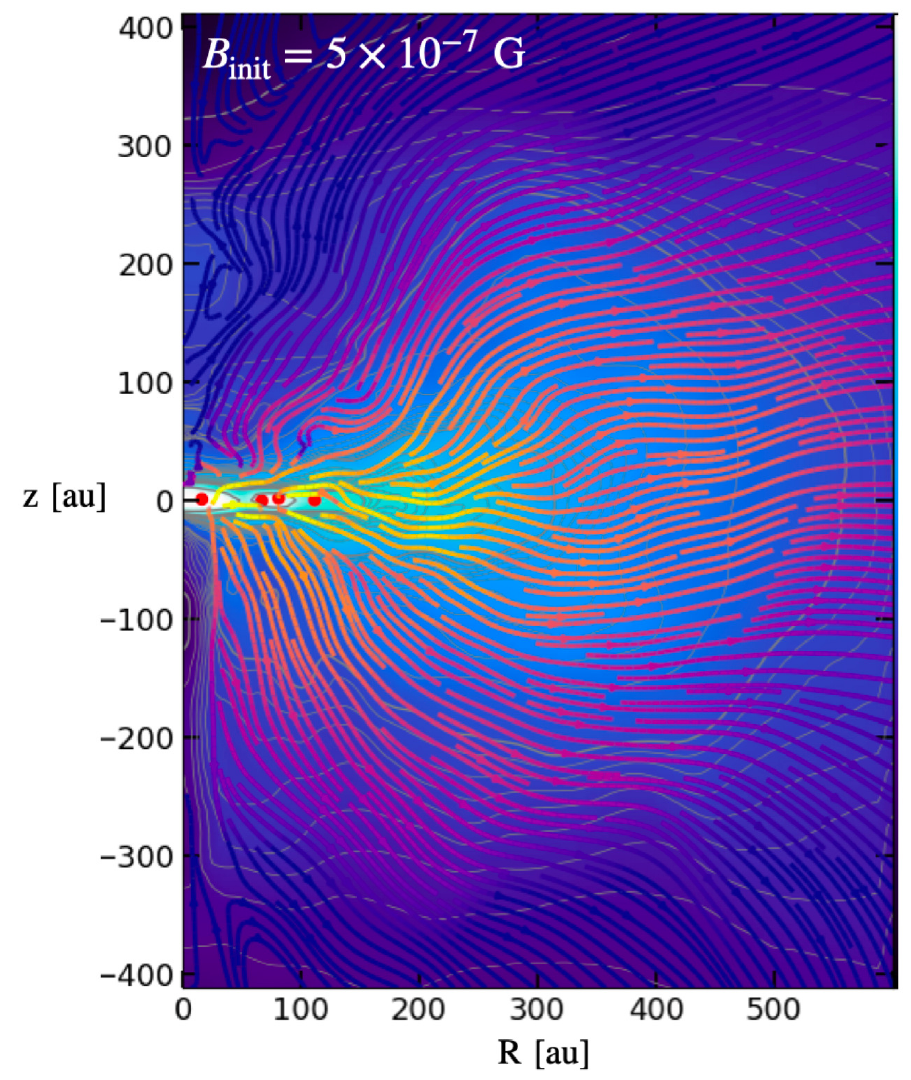} 
\end{center}
\caption{
Angular momentum transport from the circumstellar disk. Streamlines of the 2D flux ($\langle R B_{\phi} B_{R}/(4\pi) \rangle,\ \langle R B_{\phi} B_{z}/(4\pi) \rangle $) are illustrated in the $R$-$z$ plane for the strongest magnetized case ($B_{\rm init}=5\times10^{-7}\ \rm G$) at $t_{\rm p} = 800 \ \rm yr$. The color of the streamlines corresponds to the strength of the flux.
The background color and red points denote the distribution of the azimuthal average of the number density and the radial position of protostars, respectively.
}
\label{fig4.2}
\end{figure}
\begin{figure*}
\begin{center}
\includegraphics[width=160mm]{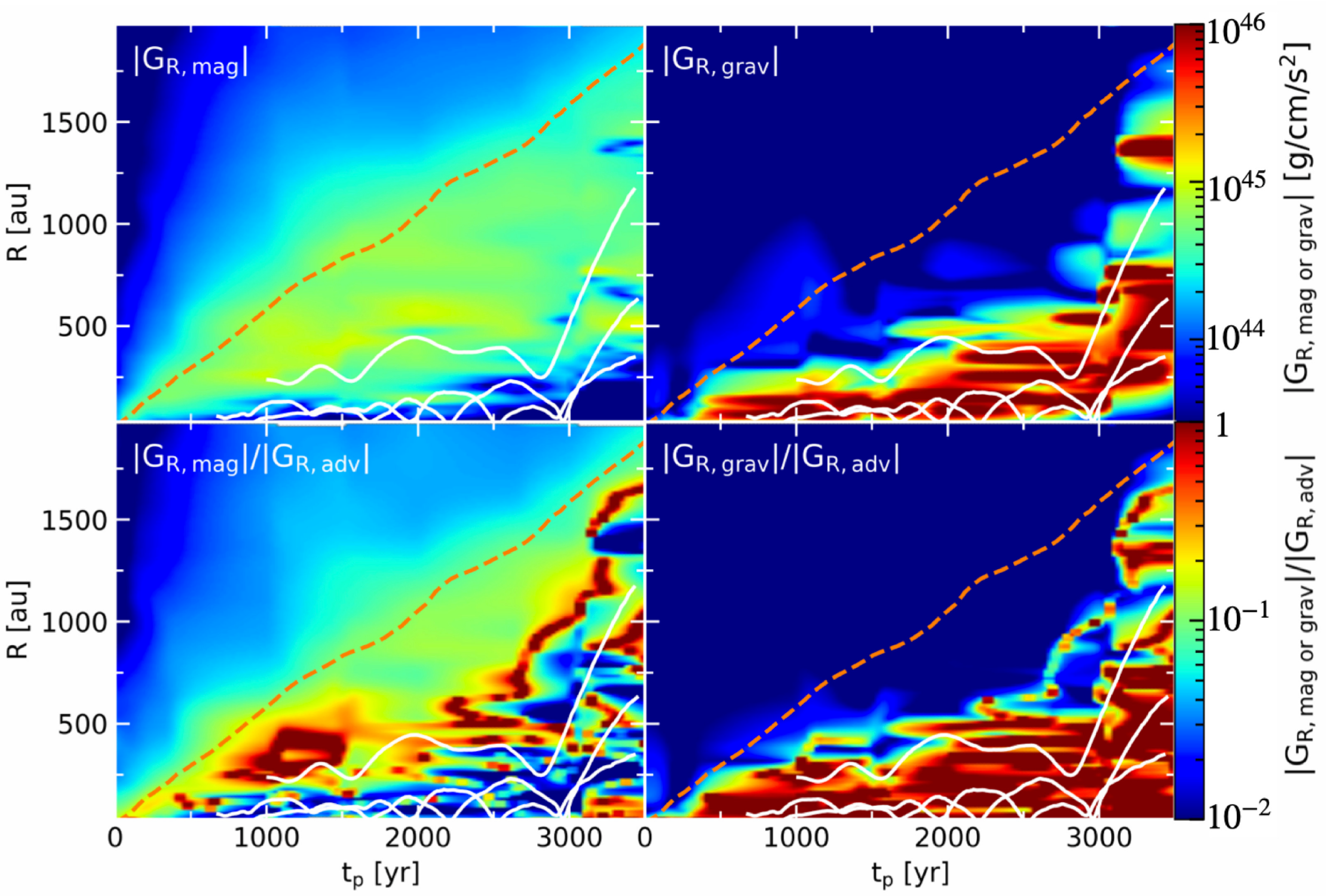} 
\end{center}
\caption{
Time evolution of the radial profiles of the radial fluxes due to magnetic torques $|G_{R,\rm mag}|$ (top left) and gravitational torques $|G_{R,\rm grav}|$ (top right), along with the ratio with respect to the advection as $|G_{R,\rm mag}|/|G_{R,\rm adv}|$ (bottom left) and $|G_{R,\rm grav}|/|G_{R,\rm adv}|$ (bottom right) in the case of $B_{\rm init}=5\times10^{-7}\ \rm G$.
The orange dashed lines and white lines indicate the disk radius and the trajectories of protostars as shown in figure \ref{fig4.4}, respectively.
 }
\label{fig4.3}
\end{figure*}
\subsubsection{Effects of magnetic torques}\label{secBtorques}
Next, we will examine the impact of angular momentum transport caused by magnetic torques.
The equation governing angular momentum conservation in cylindrical coordinates is given by 
(e.g., \cite{Joos2012}):
\begin{eqnarray}
\label{eq_AM}
\frac{\partial (\rho R v_{\phi})}{\partial t}
 &+& \nabla \cdot R \left[  \rho v_{\phi} \bm{v} + \left( P + \frac{\bm{B}^2}{8\pi} - \frac{\bm{g}^2}{8\pi G}\right)\bm{e}_{\phi} \right. \nonumber \\ 
 &-& \left. \frac{B_{\phi}}{4\pi} \bm{B} + \frac{g_{\phi}}{4\pi G} \bm{g} \right]  = 0,
\end{eqnarray}
where $\rho$, $P$, $\bm{v}$, $\bm{B}$ and $\bm{g}$ are the gas density, gas pressure, velocity, magnetic field, and gravitational acceleration, respectively.
The terms $R \rho v_{\rm \phi}\bm{v}$, $RB_{\phi}\bm{B}/(4\pi)$, and $R g_{\phi}\bm{g}/(4\pi G)$ in the equation represent the angular momentum flux due to gas advection, magnetic torques, and gravitational torques, respectively. In practice, while the advection term also includes transport due to turbulent torque, advection remains the dominant component in most cases here.

The angular momentum in the disk is primarily transported through the above three mechanisms. Angular momentum is brought into the disk region first by gas accretion from outside. This gas continues to move inward until it reaches the centrifugal radius, typically located inside the disk radius $R_{\rm disk}$ defined in this study. Consequently, angular momentum flows inward between the centrifugal radius and the disk radius due to advection. Gravitational and magnetic torques can also play a role in transporting angular momentum brought in by advection. Gravitational torques primarily arise from the asymmetric structure of the spiral arms and transport angular momentum only in the radial ($R$) direction, along with the gas material. Conversely, magnetic torques have the ability to transport angular momentum in both the radial and vertical ($z$) directions of the disk, depending on the configuration of the magnetic field. In general, angular momentum tends to be transported along the field lines.

The structure of a disk undergoes changes depending on the direction of angular momentum transport.
Vertical transport can result in the extraction of angular momentum from the disk, causing it to contract. 
While, radial transport leads to the expansion of the disk radius as outer gas receives angular momentum 
from the inner region. Consequently, the surface density in the outer part of the disk decreases, stabilizing 
it against gravitational instability.

To investigate the direction of angular momentum transport by magnetic torques, we depict the streamlines of the 2D flux $\bm{F}_{\rm mag} = (-R\langle B_{\phi} B_{R}\rangle/(4\pi) ,\ -R\langle B_{\phi} B_{z} \rangle/(4\pi) )$ in the $R$-$z$ plane for the most strongly magnetized case ($B_{\rm init}=5\times10^{-7}\ \rm G$) at $t_{\rm p} = 800 \ \rm yr$ in figure \ref{fig4.2}. Here, $\langle \cdot \rangle$ represents the volume average in the azimuthal ($\phi$) direction, and the color of the streamlines indicates the strength of the flux. The background color and red points denote the distribution of the azimuthal average of the number density and the radial position of the protostars, respectively.
From this figure, we can see that the magnetic torques primarily transport angular momentum in the radial direction. 
This implies that when turbulent magnetic fields accrete onto the disk, angular momentum cannot be extracted from the disk region. Consequently, as shown in figure \ref{fig4.1}, the disk radius tends to expand compared to the weaker magnetized cases ($B_{\rm init}\leq 10^{-7}\ \rm G$). 

To analyze the contribution of magnetic torques as angular momentum transport, 
we evaluate the strength of the angular momentum flux due to the magnetic torques $G_{R,\rm mag}$, gravitational torques $G_{R, \rm grav}$, and advection induced by the gas accretion $G_{R, \rm adv}$.
These fluxes are defined as 
\begin{equation} \label{eq_Gr_mag}
G_{R,\rm mag}(R,t) = -\int^{2\pi}_{0}\int^{h}_{-h}
R  \frac{B_{\phi}(R,\phi,z)B_{R}(R,\phi,z)}{4\pi}
R d\phi dz,
\end{equation}
\begin{equation} \label{eq_Gr_grav}
G_{R,\rm grav}(R,t) = \int^{2\pi}_{0}\int^{h}_{-h}
R  \frac{g_{\phi}(R,\phi,z)g_{R}(R,\phi,z)}{4\pi G}
R d\phi dz,
\end{equation}
\begin{equation} \label{eq_Gr_adv}
G_{R,\rm adv}(R,t) = \int^{2\pi}_{0}\int^{h}_{-h}
R  \rho v_{\phi}(R, \phi,z)v_{R}(R,\phi,z)
R d\phi dz,
\end{equation}
where $h$ is set to $300\ \rm au$, chosen to be larger than the disk height, although this choice does not significantly impact the flux strength.
\begin{figure*}
\begin{center}
\includegraphics[width=160mm]{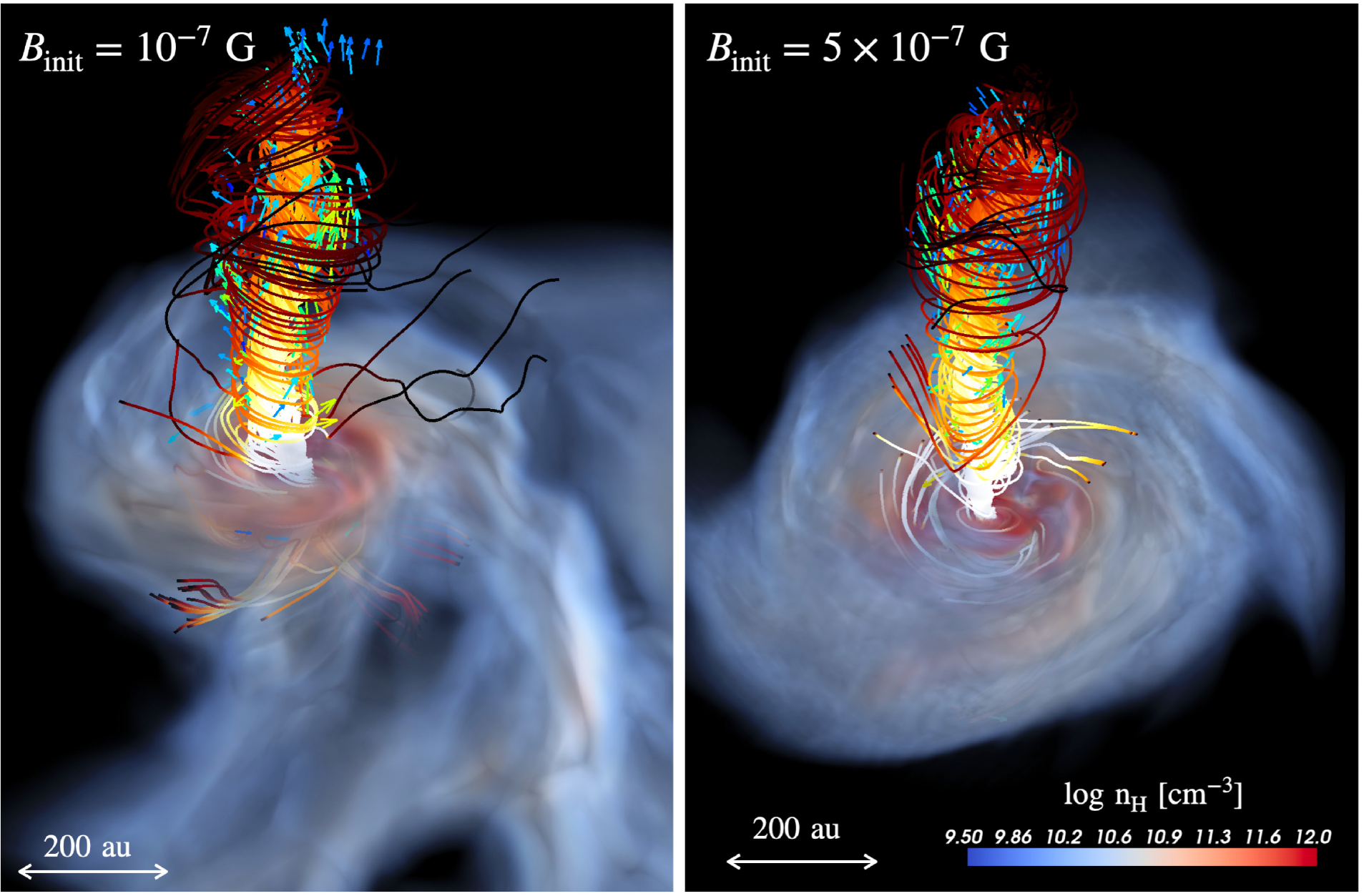} 
\end{center}
\caption{
Bird's-eye views of protostellar jets in the cases with $B_{\rm init}=10^{-7}\ \rm G$ at $t_{\rm p}=3800\ \rm yr$ (left) and $B_{\rm init}=5\times 10^{-7}\ \rm G$ at $t_{\rm p}=700\ \rm yr$ (right). 
In the case of $B_{\rm init}=10^{-7}\ \rm G$, the jet is launched from a protostar formed by the fragmentation of a spiral arm in the circum-multiple disk, which is the same 
as shown in figure \ref{fig21Bmag}. In the case of $B_{\rm init}=5\times10^{-7}\ \rm G$, the jet originates from the primary protostar.
 Solid lines depict magnetic field lines color-coded by their strength, while arrows indicate the direction of outflowing velocity fields. Gas distribution is visualized using volume rendering. 
 }
\label{fig_jet_m4}
\end{figure*}

In the top panels of figure \ref{fig4.3}, we present the temporal evolution of the radial 
profile of $|G_{R,\rm mag}|$ (left) and $|G_{R, \rm grav}|$ (right) in the case of $B_{\rm init}=5\times10^{-7}\ \rm G$.
When we compare these two panels, it becomes apparent that the flux of magnetic torques surpasses that of gravitational torques in the outer disk region. Conversely, the inner part is predominantly influenced by gravitational torques, where the spiral arms develop and the binary system is located. This region roughly corresponds to inside the centrifugal radius.

To compare these torques with advection, we present their ratios $|G_{R, \rm mag}|/|G_{R, \rm adv}|$ (left) and 
$|G_{R, \rm grav}|/|G_{R, \rm adv}|$ (right) in the bottom panel of figure \ref{fig4.3}.
In the left panel, we observe that magnetic torques are enhanced within the disk region due to the increased strength of $B_{\rm \phi}$ resulting from disk rotation. Consequently, within the disk region (below the orange dashed line), magnetic torques can transport outward $10-20 \%$ of the angular momentum carried by advection (corresponding to green or yellow regions in the bottom left panel). This leads to a decrease in surface density of the outer part of disk region, stabilizing the disk and suppressing the development of the spiral arms there. As a result, the density distribution within the disk is concentrated toward the center compared to the weaker field cases (see figure \ref{fig4.4}).
For gravitational torques (bottom right panel), they are comparable to advection in inner part.

For $B_{\rm init}=10^{-8}\ \rm G$ and $10^{-7}\ \rm G$, 
the field strength across the entire circum-multiple disk is insufficient to produce noticeable magnetic torques for stabilizing it. 
However, in some circum-stellar disks as shown in figure \ref{fig21Bmag}, the magnetic fields are amplified to approximately $\beta_{\rm p} \sim 10$, where the magnetic torques surpass gravitational torques.
\begin{figure*}
\begin{center}
\includegraphics[width=100mm]{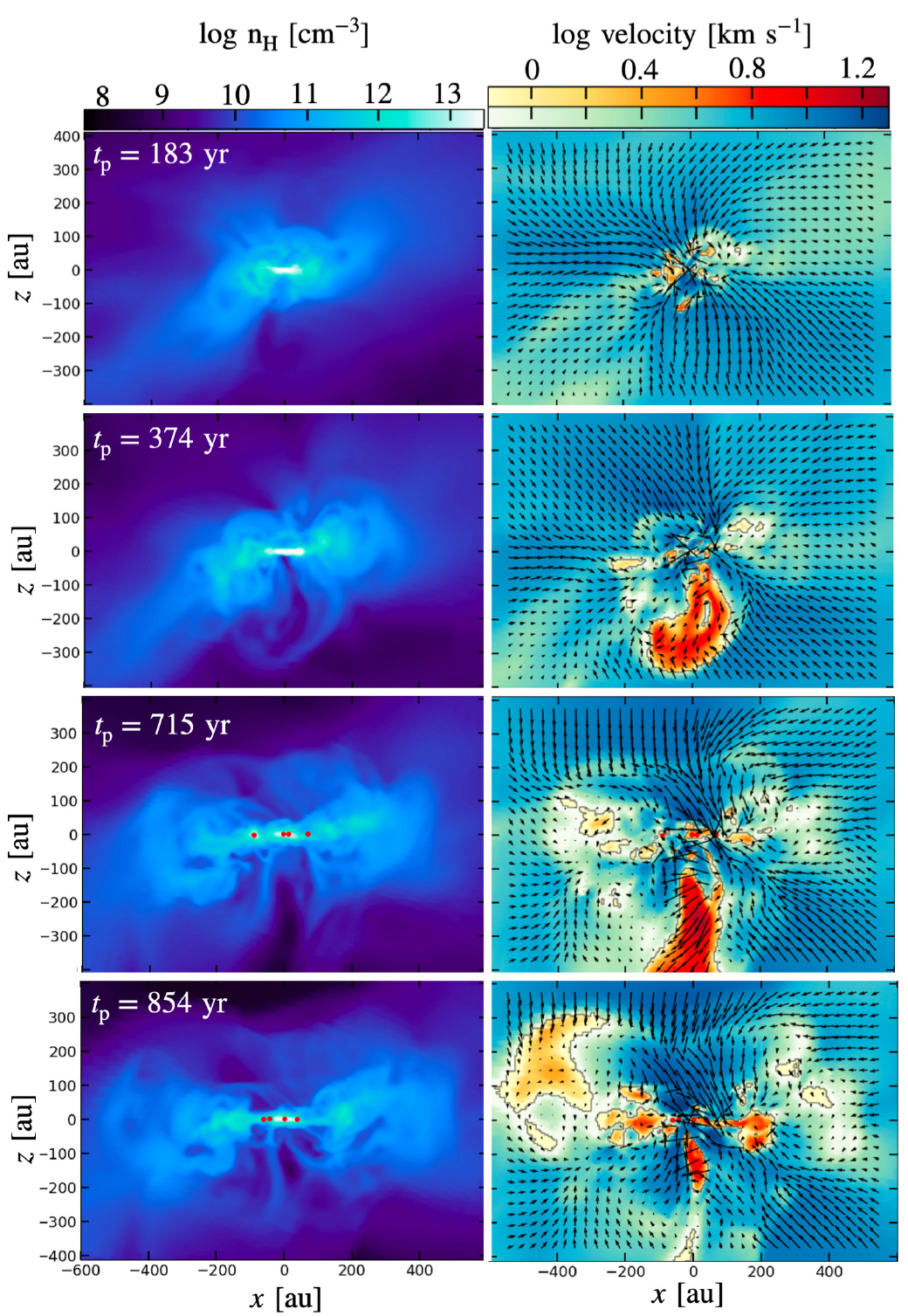} 
\end{center}
\caption{
Series of edge-on slices depicting density (left) and radial velocity (right) structures in the case of $B_{\rm init}=5\times10^{-7}\ \rm G$. The red and blue colors in the velocity panel represent outflowing and inflowing regions, respectively. Black arrows denote the direction of the velocity field sliced in the edge-on plane.
 }
\label{fig_jet_evo_m4}
\end{figure*}
\subsubsection{MHD outflows}
In general, magnetic fields can drive outflows or jets that expel gas with angular momentum, 
consequently reducing star formation efficiency. MHD outflows arise from two mechanisms: 
the magneto-centrifugal wind (e.g., \cite{Blandford_Payne1982}), driven by magnetic and 
centrifugal forces, and the magnetic pressure-driven wind (e.g., \cite{Tomisaka2002}; 
\cite{Banerjee_Pudritz2006}; \cite{Machida2008a}), driven by the gradient of toroidal 
magnetic pressure. The former type of outflows initially require coherent magnetic fields 
with strengths around $\beta_{\rm p}<1$ to permeate rotating objects like protostars or disks. 
On the other hand, for the latter type, initially uniform strong magnetic fields are not mandatory. 
This mechanism operates by twisting magnetic field lines through rotating objects, amplifying 
toroidal magnetic fields, and creating a gradient of magnetic pressure.

We observed the launching of outflows in the cases of $B_{\rm init}=10^{-7}\ \rm G$ and $5\times10^{-7}\ \rm G$, 
where magnetic fields reach equipartition during the collapse phase. In both cases, the outflows 
are driven as magnetic pressure-driven winds originating from the base of the artificial adiabatic cores 
(protostars), resembling certain types of ``protostellar jets''. The outflows blow as a pressure-driven wind rather 
than a centrifugal wind because the field strength around the protostar before the jet formation 
is not as strong as $\beta_{\rm p}\sim 1$, and the configuration is disturbed by turbulence rather 
than being coherent. The rotation of a protostar swiftly twists the magnetic field lines, generating 
a strong toroidal magnetic pressure capable of overcoming gravity and the ram pressure of gas accretion.

Figure \ref{fig_jet_m4} shows the three-dimensional view of the protostellar jets under the case of $B_{\rm init}=10^{-7}\ \rm G$ at $t_{\rm p}=3800\ \rm yr$ (left panel) and $5\times10^{-7}\ \rm G$ at $t_{\rm p}=700\ \rm yr$ (right panel). Solid lines denote magnetic field lines with their colors indicating field strength, while arrows represent the velocity field of gas being ejected outward against gravity. From these figures, we can clearly see that the fields lines inside the outflowing region are tightly twisted by the rotation of the protostar, generating a strong outward magnetic pressure gradient. This leads to a highly collimated jet structure, shaped by the pinch effect of the toroidal fields (e.g., \cite{Tomisaka2002}), forming a tower-like configuration. This type of jet resembles magnetic tower jets observed in MHD simulations around an accretion disk around BH (e.g., \cite{Kato2004}) which can be driven by various field structures. In addition, both jets exhibit a unipolar type. This result is consistent with findings from previous MHD simulations, indicating that unipolar outflows are likely to occur when the initial magnetic fields energy is weaker than the turbulent energy (e.g., \cite{Mignon2021b}; \cite{Takaishi2024}). 

In the case of $B_{\rm init}=5\times10^{-7}\ \rm G$, the protostellar jet appears from the primary protostar. Figure \ref{fig_jet_evo_m4} shows the temporal evolution of the edge-on slice of the density structure (left column) and velocity field  (right column) from the birth to extinction of the protostar jet. In the velocity map, the outflowing and inflowing region are depicted as red and blue, respectively. From this figure, we can observe that the jet begins to emerge from one side at around $t_{\rm p} \simeq 300\ \rm yr$ (second panel of figure \ref{fig_jet_evo_m4}). Although the jet maintains its collimation and continues to extend outward, its strength begins to decrease after about $720\ \rm yr$. This weakening is attributed to the high ram pressure resulting from high accretion rates. As a consequence, the jet eventually halts around $860\ \rm yr$ (bottom panel of figure \ref{fig_jet_evo_m4}), causing the blown gas to fall back toward the protostar. This phenomenon resembles the so-called 'failed outflows' observed in simulations of high-mass star formation within present-day environments (e.g., \cite{Matsushita2017}; \cite{Machida2020}). 
Consequently, the lifespan of the jet is very brief and its impact on gas ejection is minor. While some gas may be entrained by the recurrent generation and dissipation of the jet, no substantial contribution towards disk fragmentation was observed.

In the weaker field case of $B_{\rm init}=10^{-7}\ \rm G$, a protostellar jet is observed only from the protostar presented in figure \ref{fig21Bmag}. This protostar forms as a result of the fragmentation of a spiral arm within the circum-multiple disk, as previously mentioned.
Due to the combined effects of disk rotation and compression of the arm, 
the magnetic fields surrounding the protostar become intensified and coherent.
Such a magnetic field configuration facilitates the generation of strong toroidal fields through its rotation, resulting in the initiation of the protostellar jet (figure \ref{fig_jet_m4}).  However, similar to the case of $B_{\rm init}=5\times10^{-7}\ \rm G$, the lifespan of the jet is relatively short, and its impact on mass ejection remains small.
Note that numerical simulations of the present-day star formation (e.g., \cite{Federrath2014b}; \cite{Federrath2015}) have suggested that protostellar jets can influence the star formation rate and the IMF by driving turbulence within the parent gas cloud. Therefore, after longer term evolution, these jets may affect the properties of first stars.
\begin{figure}
\includegraphics[width=80mm]{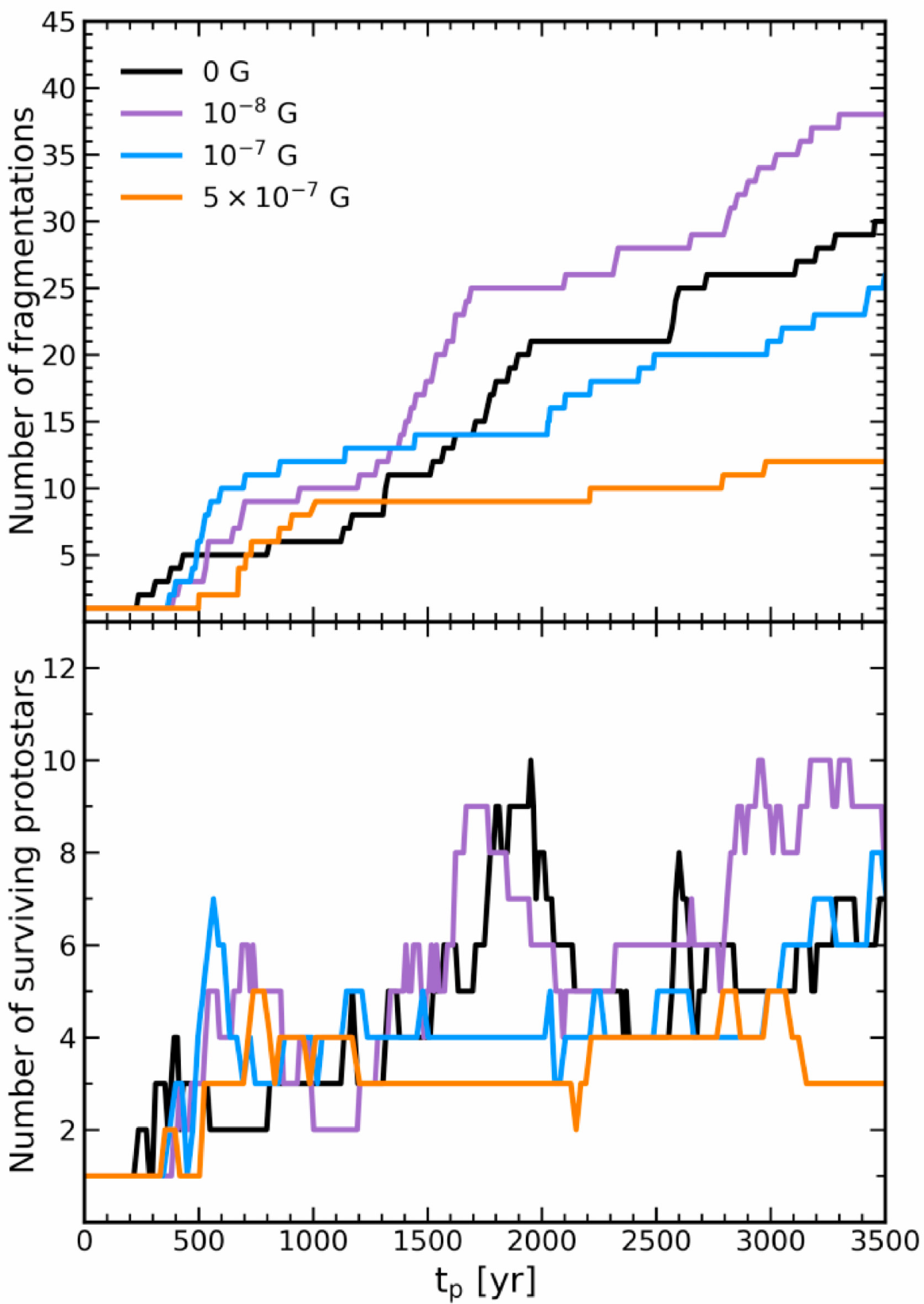} 
\caption{Time evolution of the cumulative number of fragments (top) and the number of surviving protostars 
(bottom) for four different cases: 
$B_{\rm init}=0\ \rm G$, $10^{-8}\ \rm G$, $10^{-7}\ \rm G$, and $5\times10^{-7}\ \rm G$.
}
\label{fig4.11}
\end{figure}
\subsection{The properties of multiple systems}\label{secNfrag}
Based on our analysis presented above, we confirm that both magnetic pressure and 
magnetic torques play significant roles in suppressing disk fragmentation. 
To quantitatively see the impact of magnetic effects on the fragmentation, we plot the 
time evolution of the cumulative number of fragmentation events $N_{\rm frag}$ in 
the top panel of figure \ref{fig4.11}. This figure clearly shows a trend of 
decreasing $N_{\rm frag}$ with stronger magnetization within the disk region. In 
particular, we observe a notable decrease in $N_{\rm frag}$ in the case of 
$B_{\rm init} = 5\times 10^{-7}\ \rm G$ compared to less magnetized cases 
($B_{\rm init}\leq 10^{-7}\ \rm G$). While variations in $N_{\rm frag}$ among 
simulations runs may stem from diverse realizations of the initial turbulence 
(e.g., \cite{Sharda2020}), our detailed examination in section \ref{secBeffects} 
conclusively verifies that this difference primarily originates from magnetic field 
effects.

Interestingly, in the weakly magnetized case with $B_{\rm init}=10^{-8}\ \rm G$ 
(violet line), we observe that $N_{\rm frag}$ exceeds that of the unmagnetized case 
($B_{\rm init}=0$, black line). This could be partly due to different realizations 
of turbulence, but it could also be attributed to the so-called magneto-Jeans instability, 
by which weak toroidal fields render spiral arms gravitationally unstable, leading to 
increased fragmentation (e.g., \cite{Lynden-Bell1966}; \cite{Elmegreen1987}; 
\cite{Kim_Ostriker2001}; \cite{Inoue_Yoshida2019}). Additionally, toroidal fields along the 
arms can hinder vertical gas motion, effectively trapping gas within the arms and 
potentially promoting fragmentation. Our simulations reveal that the magnetic fields within 
the arms are coherent along the arms (figure \ref{fig21Bmag}), suggesting that such magnetic 
effects are indeed plausible. 

In examining the number of surviving protostars in the bottom panel of figure \ref{fig4.11}, 
we find that the protostar count is less than half of the number of fragmentation events in 
all cases. This outcome arises primarily from the merging of multiple protostars. 
In weakly magnetized cases ($B_{\rm init}\leq 10^{-7}\ \rm G$), the protostar count exhibits episodic
oscillations. Sharp spikes in the bottom panel of figure \ref{fig4.11} correspond to circum-stellar disk 
fragmentation events. However, such protostars often migrate towards the central protostar, 
leading to frequent mergers. Nevertheless, some protostars survive and their number gradually 
increases over time, consistent with previous studies (e.g., \cite{Susa2019}).
On the other hand, in the strongly magnetized case ($B_{\rm init}=5\times10^{-7}\ \rm G$), 
such fragmentation itself is suppressed by magnetic effects. As a result, the protostar count 
remains low and relatively constant. This suggests that the influence of magnetic effects on the number 
of protostars may become more pronounced during later phases of accretion, allowing for 
clearer distinctions between them. 

Furthermore, it should be noted that in our simulations, protostars are more likely to merge 
with each other than in reality. This is because the adiabatic core is larger than the real protostar 
size. Therefore, in reality, we should expect that the protostars are not merging, but are 
instead being formed as closed binary systems or ejected out of the disk by gravitational interactions.
The difference in the number of protostars due to magnetic effects may be clearly visible in reality, 
as shown in the top panel of figure \ref{fig4.11}. In particular, the ejected protostars can be expected to 
become low-mass first stars that are observable in the present-day Universe below 
$0.8\ M_{\rm \odot}$, as accretion stops halfway through. In other words, 
strong magnetic fields amplified to about equipartition suppress the formation of low-mass stars, 
consistent with the observed fact that they have not yet been found 
(e.g., \cite{Hartwig2015}; \cite{Ishiyama2016}; \cite{Magg2018}).

The size of protostar can also influence the outflow velocity of protostellar jets. The larger protostellar radius 
in our simulations results in a shallower gravitational potential, leading to slower velocity. Conversely, 
when considering the actual radius, a stronger jet can blow, influencing the protostellar properties, 
such as their mass and spin. 

Focusing on the separation of the most massive binary systems, we observe minimal changes across 
different magnetized cases (see the trajectories of black lines in figure \ref{fig4.4}). 
In the case of $B_{\rm init}=5\times10^{-7}\ \rm G$, the binary separation remains relatively 
small until $t_{\rm p}\sim 3000\ \rm yr$, but it widens due to gravitational interactions with 
the third protostar. This suggests that in turbulent magnetic fields, the efficiency of angular 
momentum transport within the disk via magnetic torques is not sufficient to reduce the separation. 
Instead, it indicates that gravitational interactions with other protostars play a more significant 
role in determining the separation. Considering that the orbital angular momentum of the binaries comes
from the accreting angular momentum, the binary separation typically has the same order as the radius 
of circum-multiple disk.  This means that in order to effectively reduce the separation, it is necessary to
reduce the accreting angular momentum. This can be achieved by ensuring that magnetic fields reach equipartition 
level during the earlier collapse phase, facilitating the extraction of angular momentum from the envelope through 
magnetic torques. 
\section{Summary and discussion}\label{secSum}
We have performed 3D ideal MHD simulations, starting from the collapse of a turbulent primordial-gas cloud core with a density of approximately $10^{3}\ \rm cm^{-3}$, extending up to the early accretion phase characterized by frequent disk fragmentation. Our objective was to investigate the impact of amplified turbulent magnetic fields on the first star formation process, focusing specifically on disk fragmentation. We paid special attention to three major magnetic effects: magnetic pressure, magnetic torques, and MHD outflows. Through systematic exploration, we analyzed how each of these effects influences gas dynamics during the accretion phase, in order to identify the necessary conditions for magnetic fields to affect the multiplicity and binary properties of the first stars.

Our findings are summarized as follows
\begin{itemize}
\item 
Initially disturbed by turbulence before the formation of the primary protostar, the magnetic field configuration gradually transitions to a toroidally dominated one due to rotational motion within the disk. Notably, coherent toroidal fields prevail within the spiral arms. The average field strength within the disk remains relatively constant due to turbulent magnetic reconnection, which dissipates magnetic energy, suggesting that the field strength is primarily determined by amplification during the collapse phase.

\item 
When magnetic fields reach equipartition fields ($B_{\rm eq}$), where the field strength on each scale is comparable to the turbulent energy, during collapse, magnetic pressure stabilizes both the disk and spiral arms against gravitational instability, leading to fewer fragmentation events.

\item 
Magnetic torques within the disk can stabilize it by radially transporting angular momentum, similar to magnetic viscosity, if the magnetic field reaches equipartition during collapse, reducing the formation of spiral arms and fragments. However, in environments with turbulent fields persisting around the disk, magnetic torques have minimal impact on disk size or binary separation.

\item
Some protostars launch well-collimated MHD outflows, known as protostellar jets, driven by magnetic pressure winds when the magnetic field reaches equipartition during collapse. These jets, however, do not contribute significantly to disk fragmentation due to their short duration.

\item
Magnetic effects, particularly magnetic pressure and magnetic torques, impact gas dynamics during the accretion phase when magnetic fields are amplified to equipartition strength during collapse. Large-scale (around the Jeans scale) magnetic fields must be comparable to turbulence energy to influence gas dynamics.

\item 
When magnetic fields meet this criterion at the onset of the accretion phase, 
both magnetic pressure and angular momentum transport via magnetic torques 
reduce the number of fragmentations, suggesting fewer low-mass stars and a top-heavy IMF. 

\end{itemize}

In our simulations, we have varied the initial strength of magnetic fields to control 
the density at which $B_{\rm eq}$ is reached during the collapse, 
i.e., $10^{13}\ \rm cm^{-3}$ for $B_{\rm init}= 10^{-7}\ \rm G$ and $10^{11}\ \rm cm^{-3}$ 
for $B_{\rm init}= 10^{-7}\ \rm G$, respectively. This was necessary because current 
  simulations cannot accurately capture the actual amplification rate of kinematic 
  dynamo due to resolution constraints. Nonetheless, analytical considerations 
  (e.g., \cite{McKee2020}) and one-zone calculations (e.g., \cite{Schober2012a}) 
  suggest that the dynamo mechanism can amplify the seed field up to the equipartition level 
  before the formation of a cloud core of $n_{\rm H}\simeq 10^{3}-10^{4}\ \rm cm^{-3}$. 
  Hence, we expect that the conditions mentioned here are easily satisfied in realistic 
  first star-forming regions, suggesting a significant magnetic impact on circumstellar disk evolution.

As suggested in our previous study of \citet{Sadanari2023}, turbulent fields reaching equipartition 
levels can undergo a transition into a more coherent configuration. Coherent fields enhance the 
efficiency of magnetic braking, facilitating efficient extraction of angular momentum from the 
collapsing cloud. As a result, the reduction in angular momentum accretion onto the disk leads 
to shrinkage in disk radius and binary separation. This is favorable for the formation of tight 
massive binary systems, which could serve as progenitors of gravitational wave events. The disc 
also becomes gravitationally stable, thereby suppressing disk fragmentation. Consequently, the 
formation of low-mass first stars is effectively inhibited, resulting in the predominance of 
massive first stars, consistent with the absence of obseved low-mass first star in the 
present Universe. 

Furthermore, protostellar jets may persist over an extended duration, potentially exerting a substantial 
influence on the characteristics of protostars, including their mass and spin. Additionally, during 
later accretion phase, the radiative feedback from protostars becomes crucial in halting gas accretion. 
Therefore, it is imperative to perform radiative MHD simulations incorporating ionization feedback 
from protostars in future studies. 
\section*{Acknowledgments}
The authors would like to thank Drs. Gen Chiaki, Sunmyon Chon, Takashi Hosokawa, 
Ralf Klessen, Masahiro Machida, Hajime Susa and Benoit Commer\c{c}on for fruitful discussions 
and useful comments. The numerical simulations were carried out on XC50  {\tt Aterui II} in Oshu 
City at the Center for Computational Astrophysics (CfCA) of the National Astronomical Observatory of Japan, 
the Cray XC40 at Yukawa Institute for Theoretical Physics in Kyoto University, and the computer cluster 
Draco at Frontier Research Institute for Interdiskiplinary Sciences of Tohoku University. This research is 
supported by Grants-in-Aid for Scientific Research (KO: 22H00149; KS: 21K20373; TM: 18H05437, 23K03464; 
KT: 21H04487, 22K0043) from the Japan Society for the Promotion of Science.  
KES acknowledges financial support from the Graduate Program on Physics for Universe of Tohoku University. 

\bibliography{hoge}

\end{document}